\documentclass[12pt]{article}
\newlength{\abstractwidth}
\usepackage{color}
\usepackage{graphicx}

\renewcommand{\thefootnote}{\fnsymbol{footnote}}
\renewcommand{\thanks}[1]{\footnote{#1}}
\newcommand{\starttext}{
\setcounter{footnote}{0}
\renewcommand{\thefootnote}{\arabic{footnote}}}

\newcommand{\bea}{\begin{eqnarray}}
\newcommand{\eea}{\end{eqnarray}}
\newcommand{\ee}{\end{equation}}
\newcommand{\be}{\begin{equation}}

\newcommand{\sm}{\smallskip}

\def\cA{{\cal A}}
\def\cB{{\cal B}}

\def\cE{{\cal E}}

\def\cG{{\cal G}}

\def\cJ{{\cal J}}

\def\cU{{\cal U}}
\def\cV{{\cal V}}

\def\bR{{\bf R}}

\def\rt{\rightarrow}

\def\det{{\rm det}}

\def\half{ {1\over 2}}
\def\p{\partial}

\def\l({\left(}
\def\r){\right)}

\def\a{\alpha}

\def\ep{\varepsilon}

\def\l{\lambda}

\def\no{\nonumber}
\def\sm{\smallskip}

\def\Bh{\hat{B}}
\def\Th{\hat{T}}
\def\shat{\hat{s}}

\begin{document}
\starttext
\setcounter{footnote}{0}

\begin{flushright}
8 February 2012
\end{flushright}

\bigskip

\begin{center}

{\Large \bf Charge Expulsion from Black Brane Horizons,  }

\vskip 0.08in

{\Large \bf  and Holographic Quantum Criticality in the
Plane\footnote{This work was supported in part by NSF grant PHY-07-57702.}}

\vskip .5in

{\large \bf Eric D'Hoker and Per Kraus}

\vskip .2in

{ \sl Department of Physics and Astronomy }\\
{\sl University of California, Los Angeles, CA 90095, USA}\\
{\tt \small dhoker@physics.ucla.edu; pkraus@physics.ucla.edu}

\end{center}

\vskip .2in

\begin{abstract}

Quantum critical behavior in 2+1 dimensions is established via holographic methods in a
5+1-dimensional Einstein gravity theory with gauge potential form fields of rank 1 and 2. These fields
are coupled to one another via a tri-linear Chern-Simons term with strength $k$. The quantum phase
transition is physically driven by the expulsion of the electric charge from inside the black brane
horizon to the outside, where it gets carried by the gauge fields which acquire charge
thanks to the Chern-Simons interaction. At a critical value $k=k_c$, zero temperature,
and any finite value of the magnetic field, the IR behavior is governed by a near-horizon
Lifshitz geometry. The associated dynamical scaling exponent depends on the magnetic field.
For $k<k_c$, the flow towards low temperature is governed by a Reissner-Nordstrom-like black brane whose
charge and entropy density are non-vanishing at zero temperature. For $k > k_c$, the IR flow is
towards the purely magnetic brane in AdS$_6$. Its near-horizon geometry is
AdS$_4 \times \bR^2$, so that the entropy density vanishes quadratically with temperature,
and all charge is carried by the gauge fields outside of the horizon.

\end{abstract}

\newpage

%\tableofcontents

\newpage

%%%%%%%%%%%%%%%%%%%%%%%%%%%%%%%%%%%%%%%%%%%
%%%%%%%%%%%%%%%%%%%%%%%%%%%%%%%%%%%%%%%%%%%
\section{Introduction}
\setcounter{equation}{0}
\label{one}
%%%%%%%%%%%%%%%%%%%%%%%%%%%%%%%%%%%%%%%%%%%
%%%%%%%%%%%%%%%%%%%%%%%%%%%%%%%%%%%%%%%%%%%

Via the AdS/CFT correspondence, classical gravity can be used to model strongly interacting
matter at finite temperature and finite charge density.   Much activity in recent years has been devoted to
finding classical solutions to various theories of gravity whose behavior bears some relation
to interesting condensed matter systems.  To the extent that strong coupling dynamics plays
an important role in a given condensed matter system, the AdS/CFT correspondence is one
of the few computational tools available.   Holographic computations of thermodynamics and
transport are relatively easy to carry out, since they reduce to questions involving classical gravity.
See \cite{Hartnoll:2009sz,Herzog:2009xv}  for reviews.

\sm

Quantum critical systems are particularly interesting to study in this regard.
On the one hand, quantum criticality appears to be an important feature of various
real world systems~\cite{Sachdev}. On the other hand, the inherent IR universality of such quantum
critical points implies that it is meaningful to model these systems by the kinds of
supersymmetric gauge theories that naturally arise in the AdS/CFT correspondence,
even though their UV details may bear no relation to any realistic condensed matter system.

\sm

Experimentally, quantum critical points are found by tuning various control parameters,
such as a pressure or a magnetic field, and  one can attempt to map out a phase
diagram in terms of these parameters.    On the gravity side, it is possible to proceed
similarly.  One important driver of holographic quantum criticality is related to how
charge is carried by such systems; in particular, the qualitative nature of the charge
carriers can change under a variation of the control parameters, and this can signal a
quantum critical point.

\sm

In the AdS/CFT correspondence, the density of charge in the boundary theory is equal to
the amount of  bulk electric flux that pierces the boundary.  In the simplest bulk solution,
the electrically charged Reissner-Nordstrom (RN) black brane, Gauss' law implies that
these flux lines must all emanate from the horizon of the black brane, and so the charge
carriers are hidden behind the horizon.   The physical interpretation in the boundary theory
is poorly understood, partly due to the fact that this RN solution has a large ground state
entropy density.\footnote{It has been suggested, e.g. in \cite{Sachdev:2010um},
that one should think of these as
fractionalized phases of matter, in analogy with the fact that horizons are known to describe
the deconfined phases of gauge theories in the AdS/CFT correspondence.}  An alternative
scenario is for there to be a source on the right hand side of Gauss' law, in which case the
electric flux lines can be generated in the bulk as one moves out radially.  One way this can
occur is if there is a condensate of a charged scalar field in the bulk; this spontaneously
breaks the U(1) gauge symmetry and corresponds on the boundary to a holographic
superfluid phase \cite{Hartnoll:2008vx}.  A second possibility is for there to exist a sea of
charged particles in the bulk; solutions of this type have been studied in \cite{Hartnoll:2010gu}.

\sm

In the present paper, we are primarily interested in yet another possibility, in which Gauss' law is modified
by  Chern-Simons interactions. In the presence of a Chern-Simons term,
gauge fluxes can themselves carry charge, and this allows for charge to be located outside
an event horizon in the absence of explicit charged particles, and without breaking the U(1)
symmetry.   This mechanism played a key role in the magnetic quantum critical points
studied in  \cite{D'Hoker:2010rz,D'Hoker:2010ij,D'Hoker:2011xw} (see \cite{Jensen:2010vd} for
another example of magnetic quantum criticality in a holographic setting).
There, the bulk theory was five-dimensional Einstein-Maxwell-Chern-Simons
theory.  Solutions were obtained that are dual to a boundary gauge theory at finite charge
density and external magnetic field.   In the bulk, the simultaneous presence of electric and
magnetic fluxes activated the Chern-Simons term, allowing bulk charge to be carried in the
manner described above.  The strength of this effect was varied by changing either the
Chern-Simons level $k$ or the value of the magnetic field $B$, and a rich phase structure
was uncovered.  Most interestingly, for sufficiently large $k$,  a magnetically tuned quantum
critical point with nontrivial scaling properties was discovered, resembling real world systems
in a number of ways \cite{D'Hoker:2010rz}.

\sm

The critical theories studied in   \cite{D'Hoker:2010rz,D'Hoker:2010ij}
are effectively 1+1 dimensional: while the underlying gauge theory is defined in D=3+1,
at the critical point massless propagation takes place along a single spatial direction.
In the bulk, this was seen from the existence of a near-horizon geometry of the form
WAdS$_3 \times \bR^2$, where WAdS$_3$ is a ``null warped" deformation of AdS$_3$ space
\cite{Anninos:2008fx,Compere:2009zj}.
The $\bR^2$ was supported by the nonzero magnetic field.  While quantum criticality in
1+1 dimensions is certainly important, it is also interesting to generalize to higher dimensions.
Particular interest attaches to the case of 2+1 dimensions, which in the real world is relevant
to various layered materials such as the cuprates. Our goal in the present work is thus to find
a magnetic quantum critical point in 2+1 dimensions, driven by the presence of Chern-Simons terms.

\subsection{Overview of the set-up}

The simplest setup exhibiting the desired behavior appears to be six-dimensional gravity
coupled to a one-form potential $A$ and a two-form potential $C$.   Assuming gauge
invariance  under $ A \rt A+  d\Lambda_A$ and $C \rt C+ d\Lambda_C$, the most
general two-derivative action is,
\bea
\label{3aa1}
S = - { 1 \over 16 \pi G_6} \int d^6x \sqrt{g}
\left ( R - {20 \over L^2} + F^{MN} F_{MN} + { 1 \over 3} G^{MNP} G_{MNP} \right )
+  S_{\rm CS} + S_{\rm bndy}
\eea
with the Chern-Simons (CS) term $S_{\rm CS}$ given by,
\bea
\label{3aa2}
S_{\rm CS} = {k \over 4 \pi G_6} \int C \wedge F \wedge F
\eea
Here $F=dA$ and $G=dC$  are the  two-form and three-form field strengths respectively.
While such an action can plausibly arise within string theory, for the time being we will regard
this as the result of a bottom-up construction.\footnote{A related theory with two-form and
three-form potentials and Chern-Simons terms is obtained from massive IIA supergravity
\cite{Romans:1985tw}, and admits solutions with a four-dimensional Lifshitz factor
\cite{Gregory:2010gx,Braviner:2011kz}, similar to what we find below.}    Although the dual
boundary theory is defined in 4+1
dimensions,\footnote{CFTs in 4+1 dimensions are rare, but not unheard of \cite{Seiberg:1996bd}.} as discussed above our primary interest is in finding solutions corresponding to
quantum criticality in 2+1 dimensions. This is achieved by introducing a magnetic field, which
acts to freeze out two of the spatial dimensions in the IR.

\sm

To turn on a nonzero charge density $\rho$ and magnetic field $B$
we consider a two-form field strength of the form
\bea
\label{3ea1}
F =  E(r) \, dr \wedge dt +  B dx^3 \wedge dx^4
\eea
The value of  the charge density $\rho$ is read off from the asymptotic behavior of $E$,
namely $\rho \sim r^4 E(r)$ as $r \to \infty$.
The CS term   induces a nonzero three-form field strength,
\bea
\label{3eb1}
G =   G_1 (r) \, dr \wedge dx^1 \wedge dx^2
\eea
The full solution is taken to be translationally invariant along all boundary directions, as well
as rotationally invariant in the $x^{1,2}$ and $x^{3,4}$ planes.  The form-field equations of
motion exhibit a source in Gauss' law due to the CS term,
\bea
\label{3aa7}
d*F - 2k F \wedge G & = & 0
\no \\
d *G + k F \wedge F & = & 0
\eea
In particular, the product $B G_1$ acts as a charge density sourcing $E$.

\subsection{Summary of the results}

Given this setup, we can try to characterize the zero temperature solutions of this theory
as a function of the charge density $\rho$, magnetic field $B$, and Chern-Simons level $k$.
Actually, due to scale invariance there are really only two parameters to vary: the dimensionless
parameter $k$ and the dimensionless ratio,
\bea
\hat{B} = {B\over \sqrt{\rho} }
\eea
One way to summarize our results is to give the low temperature behavior of the entropy
density $s$ for various values of $\hat{B}$ and $k$. Let us first note two special cases that
are easy to understand.  For $\hat{B}=0$ the solution is just the electric RN black brane, with a
finite entropy density at zero temperature.   The other extreme is $\hat{B} = \infty$, attained
with $E=\rho=0$.  In this case we find a six-dimensional version of the magnetic brane solution
studied in \cite{D'Hoker:2009mm} for five dimensions.   The solution develops a near-horizon
AdS$_4 \times \bR^2$ geometry which connects to the asymptotic AdS$_6$.
In both of these cases the three-form field strength $G$ vanishes identically,
and the Chern-Simons term plays no role.

\sm

For finite nonzero values of $\hat{B}$ we find that the qualitative behavior depends on the
magnitude of $k$ relative to its critical value $k_c = 1/\sqrt{3}$.   We discuss the three cases in turn.

\subsubsection{The case $k<k_c$}

Solutions in this regime can be thought of as  deformed versions of the purely electric RN solution.
At $\Bh =0$, we have precisely the RN solution, as $G_1$ vanishes  and the CS term plays no role.
At zero temperature,  the near-horizon region becomes AdS$_2 \times \bR^4$, and the zero
temperature entropy density is proportional to the charge density:  $s \propto \rho$.
For this RN solution, all of the charge is hidden behind the horizon.

\sm

\begin{figure}[htb]
\begin{centering}
\includegraphics[scale=0.8]{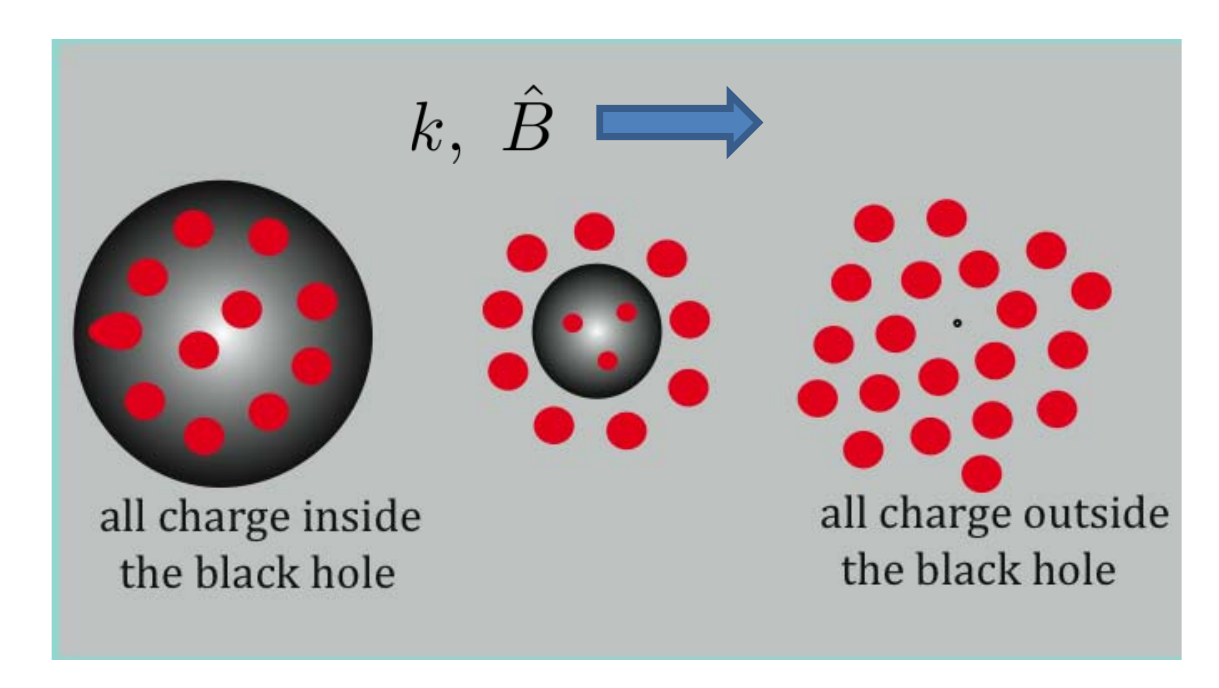}
\caption{As $k$ and $\hat B$ are increased (indicated by the blue arrow), the charge  inside the horizon of the
black hole gets expelled and carried by the gauge fields outside the horizon. }
\label{fig1}
\end{centering}
\end{figure}

Turning on nonzero $\Bh$ activates $G_1$ and the CS term.   At zero temperature the near-horizon region again exhibits an AdS$_2 \times \bR^4$, but now only a finite fraction of the charge
is hidden behind the horizon, with the remainder carried outside the horizon by the flux term $BG_1$.
As $k$  and/or $\hat B$ is increased, this effect becomes more pronounced, with more charge
transferred from behind the horizon to outside, as illustrated in figure \ref{fig1}. The transfer becomes complete as $k$ reaches its
critical value $k_c$, at which point there is a quantum phase transition. For $k<k_c$, all of these solutions have a nonzero
ground state entropy density.

\subsubsection{The case $k>k_c$}

Here the solutions are generalizations of the purely magnetic brane solution obtained for $E=G_1=0$.
The purely magnetic solution interpolates between a near-horizon AdS$_4 \times \bR^2$ region and
AdS$_6$ asymptotically, and is the analog of the 5D magnetic brane solution studied in
\cite{D'Hoker:2009mm}. The low temperature entropy density  behaves like that of a D=2+1 Lorentz
invariant CFT, and exhibits the expected $s \propto T^2$.  At small but nonzero
temperature, the near-horizon region
has AdS$_4$ replaced by an AdS$_4$ Schwarzschild black brane.

\sm

Turning on nonzero charge density changes the details, but the qualitative picture remains.
The zero temperature solutions always exhibit a near-horizon AdS$_4 \times \bR^2$ region and the
entropy density is quadratic in the temperature.   The dependence on $\Bh$ and $k$ can be seen
in the behavior of the prefactor,
\bea
\shat \sim A(k,\Bh) \, \Th^2
\eea
where we now work in terms of dimensionless quantities, referred to with hats, and to be defined
precisely in (\ref{3g5}).  It is interesting to explore the behavior of $A(k,\Bh)$ as $\Bh$ tends to 0,
since precisely at $\Bh=0$ (obtained for $B=0$ and $\rho \neq 0$) we arrive back at the purely electric
RN solution with its finite  entropy density ground state.  The coefficient $A(k,\Bh)$ must therefore
exhibit a singularity, and indeed we find the following behavior, as $\hat B \to 0$,
\bea A(k,\Bh) \sim   c(k) \exp \left \{ {d(k) \over   \Bh^2} \right \}
\eea
for real positive functions $c(k)$ and  $d(k)$.  We present clear numerical evidence for the simple but distinctive
analytical behavior of $A(k,\hat{B})$ (and for $d(k)$ as explained below), even though at present the corresponding
analytical derivations are not available.

%This result has been extracted numerically.

\sm

For $k> k_c$ we therefore find that the ground state entropy density of the purely electric RN
solution is removed by the presence of any nonzero magnetic field $B$.    For any finite $B$,
the entropy density will always go to zero quadratically in $T$ as $T\rt 0$.  The finite ground
state entropy density of the RN solution thus requires a fine tuning, since it is removed by turning
on an arbitrarily small magnetic field.  It is instructive to compare this to the behavior of the
analogous charged magnetic brane solutions in AdS$_5$  studied in \cite{D'Hoker:2009bc}. There,
we conjectured, based on preliminary numerical analysis, that the AdS$_5$ RN ground state entropy
density would similarly be removed by the presence of any finite magnetic field.  However,
subsequent high-precision numerical work \cite{D'Hoker:2010rz} and analytical
calculations \cite{D'Hoker:2010ij} revealed that the magnetic field needed to be larger than a value
$B_c$ to achieve this.  So in this sense the effect of the magnetic
field appears to be stronger in the AdS$_6$ case as compared to AdS$_5$.

\sm

The function $A(k,\Bh)$ is also singular as $k$ approaches $k_c$ from above. Numerical
analysis shows that the coefficient $d(k)$ behaves as
\bea
d(k) \sim { d_0 \over k^2 - k_c^2}
\eea
to remarkably good accuracy in the range $k_c < k < 1.2$. Here, $d_0$ is a constant of order 1.
This again signals a quantum phase transition at $k_c$, to which we now turn.

\subsubsection{The case $k=k_c$}

Numerical analysis at $k=k_c$ indicates that the low temperature entropy density  scales to zero
as a $\Bh$-dependent power of temperature,
\bea
s\propto T^{2 / z}
\hskip 1in
z=z(\Bh) \rt  \left\{\matrix{\infty & \hbox{as} & \Bh \rt 0 \cr && \cr 1 & \hbox{as} & \Bh \rt \infty}  \right.
\eea
A plot of $z(\hat{B})$ will be shown in figure \ref{critexp}.
Such a behavior is characteristic of a 2+1 dimensional scale invariant theory with dynamical critical
exponent $z=z(\Bh)$.   Indeed, precisely at $k=k_c$ the field equations admit a new near-horizon geometry of the form Lif$_4 \times \bR^2$, where Lif$_4$ is a 3+1 dimensional
Lifshitz solution \cite{Kachru:2008yh} with dynamical exponent $z$.   Explicitly,
\bea
\label{nhlif}
ds^2 &=& {dr^2 \over u_0 r^2} - u_0 r^2 dt^2 + r^{2/z}\Big( (dx^1)^2 + (dx^2)^2\Big) + \Big( (dx^3)^2 + (dx^4)^2\Big)
\no \\
F & = & q_0 dr \wedge dt + b_0 dx^3 \wedge dx^4
\no \\
G & = & -{q_0 \over kb_0 z} r^{2/z-1}dr \wedge dx^1 \wedge dx^2
\eea
In terms of $z$ restricted to the range $1 \leq z \leq \infty$, as is required for all fields to be real valued,
the values of $q_0$ and $b_0$ are fixed as
\bea
q_0^2 =  {10 z(z-1) \over (z+2)^2}
\hskip 1in
b_0^2 ={ 30z \over (z+2)^2}
\eea
The parameters $q_0$ and $b_0$ represent the near-horizon value of the electric charge and magnetic field
respectively, and are related to asymptotic quantities measured at the AdS$_6$ boundary by the full
interpolating solution.  There is thus a nontrivial expression for $\Bh$ in terms of $z$, or equivalently
$z(\Bh)$, whose form is only known numerically; see figure \ref{critexp} below.

\sm

Numerical analysis shows that as the temperature is lowered at $k=k_c$ a near-horizon region of the form (\ref{nhlif})  indeed develops and controls the low temperature thermodynamics. Also, as $\Bh$ varies from $0$ to $\infty$, the dynamical exponent $z$ smoothly varies between $\infty$ and $1$, so that the family of Lifshitz geometries interpolates between AdS$_2 \times \bR^4$ and AdS$_4\times \bR^2$.

\sm

\begin{figure}
\centering
\begin{minipage}[c]{0.5\linewidth}
\centering \includegraphics[width=3.4in]{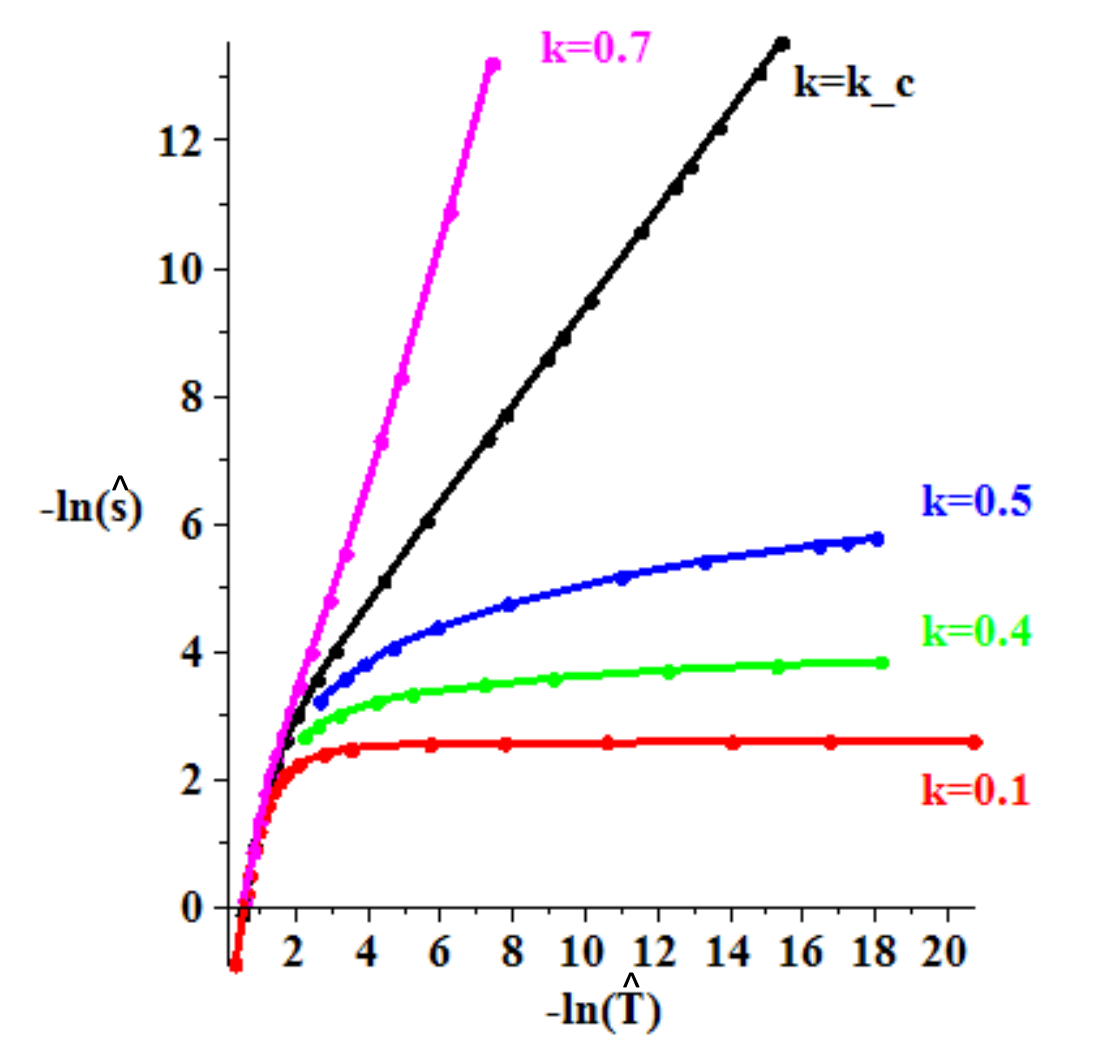}
\end{minipage}%
\begin{minipage}[c]{0.5\linewidth}
\centering \includegraphics[width=3.3in]{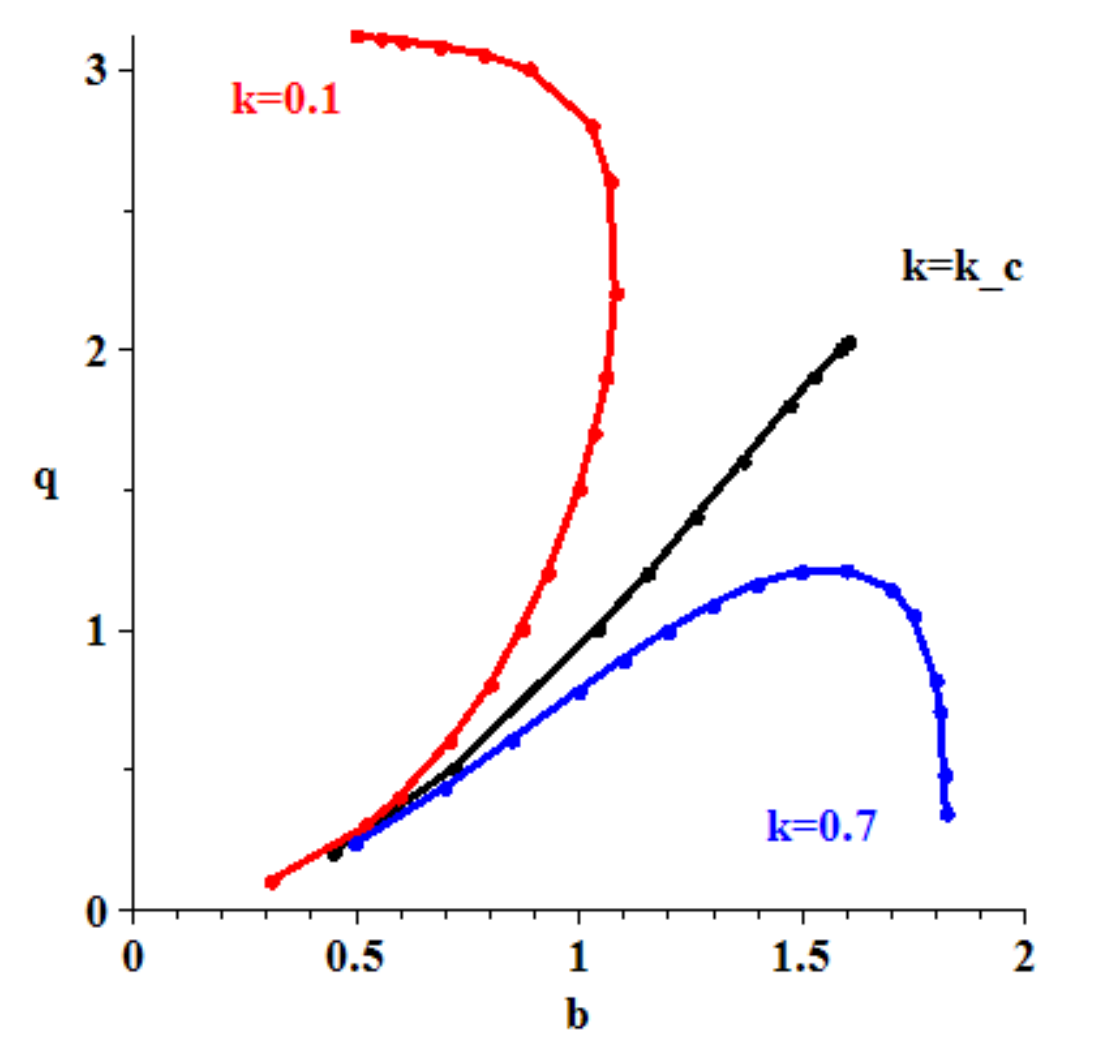}
\end{minipage}
\caption{Left panel: Entropy density $\hat s$ versus temperature $\hat T$ for selected values of $k$.
Right panel: Flows towards low temperature of the horizon data $q$ and $b$, for fixed $\hat B=1$,
for $k=0.1$ (red dots), $k=k_c$ (black dots), and $k=0.7$ (magenta dots). }
\label{fig2}
\end{figure}

In figure \ref{fig2}, we display representative numerical data that illustrate the points made above.  In the left panel $(b,q)$ denote the values of the electric and magnetic fields at the horizon (in a particular coordinate system described in the text).  By varying the temperature for a fixed value of $\hat{B}$ we obtain curves in $(b,q)$ space.  The curves emanate from the origin, which corresponds to the high temperature regime. As the temperature is lowered, the curves either flow to the electric fixed point at $b=0$ for $k<k_c$; to the magnetic fixed point at $q=0$ for $k>k_c$; or to a point on a critical curve with nonzero $q$ and $b$ for $k=k_c$.   In the right panel we show the corresponding behavior of the entropy density as a function of temperature.   As $T \rt 0$, the entropy density either goes to a finite value ($k<k_c$); to zero quadratically in temperature ($k>k_c$); or to zero with a $\hat{B}$ dependent power between $0$ and $2$ ($k=k_c$).

\sm

\begin{figure}[htb]
\begin{centering}
\includegraphics[scale=0.65]{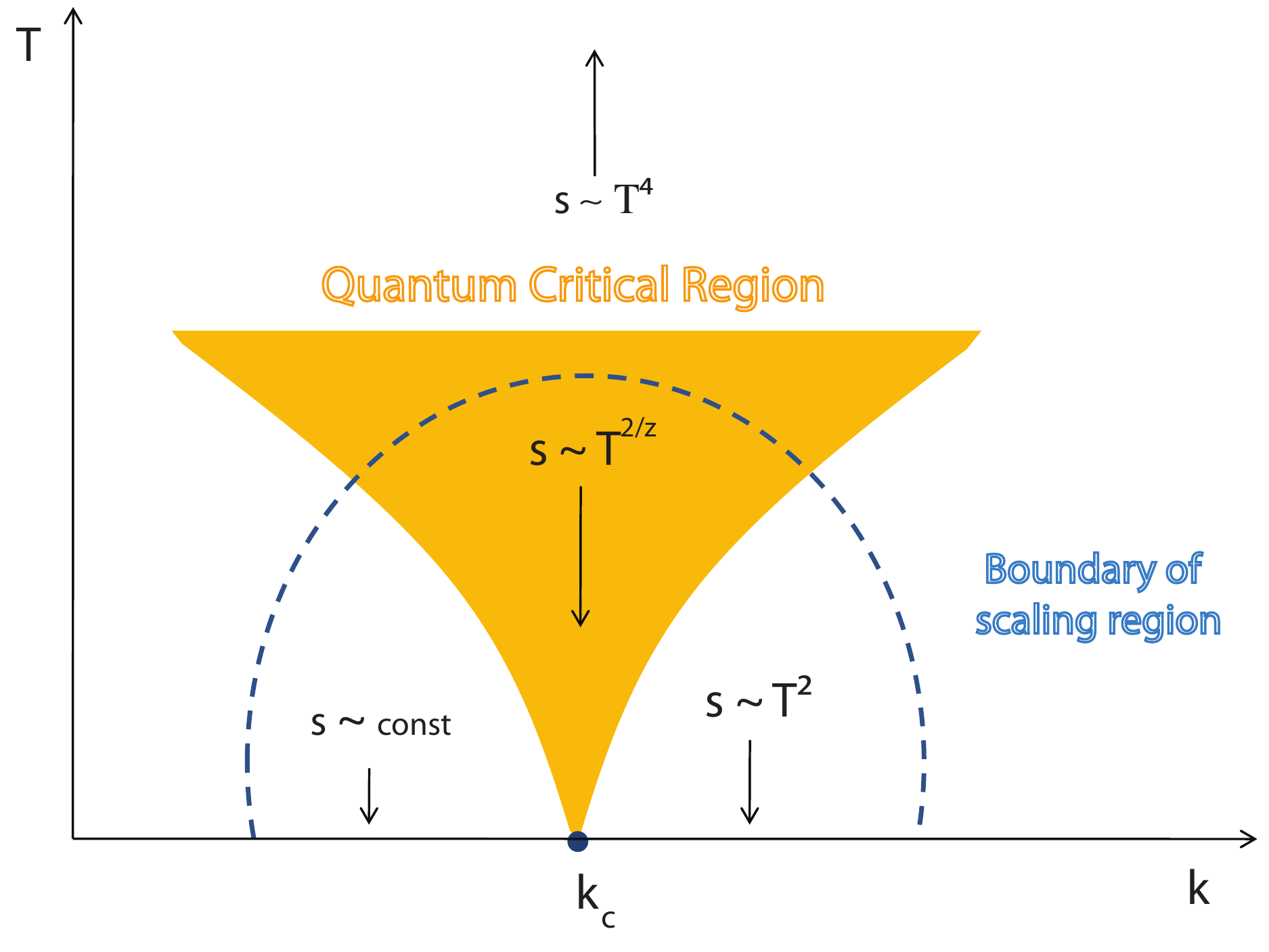}
\caption{Phase diagram of entropy density $s$ versus $k$ for all finite $\hat B$.}
\label{fig3}
\end{centering}
\end{figure}

Examining the overall picture, summarized in figure  \ref{fig3},
the story is quite similar to that in the AdS$_5$ case,
except that the roles of $k$ and $B$ are, in a sense, interchanged.  In the AdS$_5$ case a critical
point was reached by tuning $B$ to $B_c$, and at the critical point the dynamical  exponent
$z$ depended on $k$.    But in the AdS$_6$ case studied here, the critical point is reached by
tuning $k$, and at the critical point the dynamical exponent varies with $B$.    The other major
distinction is, of course, that for AdS$_5$ the critical theory was 1+1 dimensional, while in the
AdS$_6$ case it is 2+1 dimensional. 

A model with related behavior was studied in \cite{Hartnoll:2011pp}.  There also, transitions were found between finite and zero entropy density states at zero temperature, mediated by the appearance of a Lifshitz near horizon geometry.

\subsection{Organization}

The remainder of this paper is organized as follows. In section \ref{two}, we present a general discussion
of candidate theories for holographic quantum critical behavior in 2+1 dimensions, built on Einstein
gravity plus gauge fields of various ranks. The theory of gauge field potentials of rank 1 and 2 in six
space-time dimensions with a Chern-Simons coupling, considered in this paper, emerges as one of the
simplest such models. In section \ref{three}, a detailed discussion of the action, field equations, symmetries,
suitable Ansatz, and reduced field equations for the study of thermodynamic quantities is
presented. In section \ref{four}, the special solutions corresponding to the well-known AdS$_6$
Reissner-Nordstrom black brane, and the novel purely magnetic brane in AdS$_6$ are derived.
The critical value of $k=k_c$ is inferred from the behavior of small fluctuations around the purely
magnetic brane. The existence of a Lifshitz near-horizon geometry at $k=k_c$ is shown in section
\ref{five}, its extension to a full fledged six-dimensional solution is derived numerically, while an
analytical derivation of the small fluctuation spectrum around Lifshitz is deferred to Appendix \ref{AppA}.
In section \ref{six}, a detailed discussion is presented of our numerical results,
and their matching with analytical calculations when available (with further details explained in Appendix B).
Finally, the quantum critical behavior of the system is discussed
in section \ref{seven}.

%\newpage

%%%%%%%%%%%%%%%%%%%%%%%%%%%%%%%%%%%%%%%%%%%
%%%%%%%%%%%%%%%%%%%%%%%%%%%%%%%%%%%%%%%%%%%
\section{Candidates with $p$-form fields}
\setcounter{equation}{0}
\label{two}
%%%%%%%%%%%%%%%%%%%%%%%%%%%%%%%%%%%%%%%%%%%
%%%%%%%%%%%%%%%%%%%%%%%%%%%%%%%%%%%%%%%%%%%

The holographic model for quantum critical behavior in 1+1 dimensions based on
4+1-dimensional Einstein-Maxwell-Chern-Simons \cite{D'Hoker:2010rz,D'Hoker:2010ij} draws on a
number of fundamental dynamical properties.
\begin{enumerate}
\item The presence of a constant background magnetic field induces an IR flow towards a
lower-dimensional AdS theory;
\item The gauge field can carry charge thanks to the Chern-Simons interaction. This mechanism
permits charge to be expelled from the inside of the black brane horizon to the outside where it is
carried by the gauge field.
\item Tri-linearity of the Chern-Simons action supplies the IR theory with an effective
Chern-Simons theory for the low-energy fields.
\end{enumerate}
The goal of this paper is to generalize and extend this construction to space-times where
quantum criticality takes place in field theories of dimension higher than 1+1. To do so,
we shall want to respect the basic tenets above. Given these assumptions, which
theories can we construct, while maintaining the basic assumption that no scalar fields be present?

\subsection{Gravity models with a single form-field}

The Einstein-Maxwell-Chern-Simons theory of \cite{D'Hoker:2010rz,D'Hoker:2010ij}, formulated in terms of
a single 1-form field potential, admits a straightforward generalization to a theory with
a single $p$-form field $C$ whose field strength form is denoted by $G=dC$. The corresponding
action is given by,
\bea
\label{2a1}
S = - { 1 \over 16 \pi G_D} \int d^Dx \sqrt{g}
\left ( R - \Lambda  +  | G |^2 \right )
+  S_{\rm CS}
\eea
where $\Lambda$ is the cosmological constant,  $S_{\rm CS}$ the Chern-Simons term given by,
\bea
\label{2a2}
S_{\rm CS} = {k \over 4 \pi G_D} \int C \wedge G \wedge G
\eea
and the boundary action has been suppressed.
Matching of dimensions requires $D=3p+2$. The Einstein-Maxwell-Chern-Simons
theory of \cite{D'Hoker:2010rz,D'Hoker:2010ij} corresponds to $p=1$. When $p>1$, we assume that
$G$ has a boundary magnetic field that generates a flow from AdS$_D$=AdS$_{3p+2}$ in the UV
to AdS$_{D-p-1}$=AdS$_{2p+1}$ in the IR. For $p=2$, we have $D=8$ and a flow towards AdS$_5$,
while for $p=3$, we have $D=11$ and flow towards AdS$_7$. The latter case might be
of special interest as arising from M-theory on a $S^4$ sphere with magnetic field.
All of these schemes may be interesting, but still they lack the presence of the
fundamental electro-magnetic field.

\subsection{Gravity models with two form-fields}

Gravity models with two different form potential fields $A$ and $C$, respectively
of ranks $p'$ and $p$, and field strengths $F=dA$ and $G=dC$, offer a further extension
of the Einstein-Maxwell-Chern-Simons models of
\cite{D'Hoker:2010rz,D'Hoker:2010ij}.\footnote{We will be assuming invariance (up to boundary
terms) of the action under the gauge invariance $A\rt A + d\Lambda_A$, $C\rt C+ d\Lambda_C$,
although further generalizations could be envisaged as well.}  The general action is as follows,
\bea
\label{2b1}
S = - { 1 \over 16 \pi G_D} \int d^Dx \sqrt{g}
\left ( R - \Lambda  +  | F |^2  +  | G |^2 \right )
+  S_{\rm CS}
\eea
where $\Lambda$ is the cosmological constant.    All possible tri-linear Chern-Simons terms are,
\bea
\label{2b2}
S_{\rm CS} ^{AAA} & = &  {k_1 \over 4 \pi G_D} \int A \wedge F \wedge F \hskip 1in D=3p'+2
\no \\
S_{\rm CS} ^{AAC} & = &  {k_2 \over 4 \pi G_D} \int A \wedge F \wedge G \hskip 1in D=2p'+p+2
\no \\
S_{\rm CS} ^{ACC} & = &  {k_3 \over 4 \pi G_D} \int A \wedge G \wedge G \hskip 1in D=p'+2p+2
\no \\
S_{\rm CS} ^{CCC} & = &  {k_4 \over 4 \pi G_D} \int C \wedge G \wedge G \hskip 1in D=3p+2
\eea
In the presence of two different form fields of the same rank $p=p'$, all of the above Chern-Simons
terms are allowed for $D=3p+2$, and the theory is characterized by 4 independent dimensionless
couplings $k_1, k_2, k_3, k_4$. On the other hand, if $p\not = p'$, and we take $p '< p$, then
for given $D$ only a single one of the above Chern-Simons couplings can occur.

\sm

Assuming that one of the potentials $A$ is electro-magnetic, we set $p'=1$, and $p>1$.
For $p=1$ we recover the $D=5$ theory considered in \cite{D'Hoker:2010rz,D'Hoker:2010ij}
except now with two gauge fields and a mixed Chern-Simons term; theories of this type were
studied in \cite{Almuhairi:2010rb,Almuhairi:2011ws}.  For $D>5$, each value of $p \geq 2$ yields
a unique theory with Chern-Simons term $S_{CS}^{AAC}$ in dimension $D=4+p$, and a
unique theory with Chern-Simons term $S_{CS}^{ACC}$ with dimension $D=3+2p$.
The next simplest case has $p=2$, $D=6$ and Chern-Simons term $S_{CS}^{AAC}$
and is the one considered here. Another interesting case is $p=2$, $D=7$ and
Chern-Simons term $S_{CS}^{ACC}$.

%\newpage

%%%%%%%%%%%%%%%%%%%%%%%%%%%%%%%%%%%%%%%%%%%
%%%%%%%%%%%%%%%%%%%%%%%%%%%%%%%%%%%%%%%%%%%
\section{The Einstein-Maxwell-two-form-field system}
\setcounter{equation}{0}
\label{three}
%%%%%%%%%%%%%%%%%%%%%%%%%%%%%%%%%%%%%%%%%%%
%%%%%%%%%%%%%%%%%%%%%%%%%%%%%%%%%%%%%%%%%%%

In this section, we present the Einstein-Maxwell-two-form-field system in
detail, including the action and field equations. To study thermodynamics in the presence
of a uniform background magnetic field and electric charge density, we focus on solutions
which are invariant under translations, and certain rotations, of the boundary theory.
We obtain the corresponding Ans\"atze and reduced field  equations.

\subsection{Field contents and field equations}

The fundamental fields of the 6-dimensional theory are: the metric $g_{MN}$, the Maxwell field
$A_M$, and the 2-form field $C_{MN}=-C_{NM}$.  The action consists of the Einstein-Hilbert term,
standard kinetic terms for the fields $A$ and $C$, and the only possible Chern-Simons term
afforded by this structure,\footnote{Throughout,
Einstein indices range over $M,N,P=0,1,2,3,4,5$, and we define $g = - \det (g_{MN})$.
Isolating the holographic radial direction $r$, and the time
direction $0$, we adopt the following convention for the orientation, $\ep _{r01234} =1$.
Expressed in Einstein indices, the dual $\star G$ of a $p$-form $G$ with components
$G_{M_1 \cdots M_p}$, is defined by  $(*G) _{N_1 \cdots N_{6-p}} = {\sqrt{g} \over p!}
\ep _{M_1 \cdots M_p N_1 \cdots N_{6-p}} G^{M_1 \cdots M_p}$.}
\bea
\label{3a1}
S = - { 1 \over 16 \pi G_6} \int d^6x \sqrt{g}
\left ( R - {20 \over L^2} + F^{MN} F_{MN} + { 1 \over 3} G^{MNP} G_{MNP} \right )
+  S_{\rm CS} + S_{\rm bndy}
\eea
with the Chern-Simons term $S_{\rm CS}$ given by,
\bea
\label{3a2}
S_{\rm CS} = {k \over 4 \pi G_6} \int C \wedge F \wedge F
\eea
The explicit form of the boundary action $S_{\rm bndy}$ will not be needed.
Here,  $G_6$ is Newton's constant in 6 dimensions, $L$ sets the scale for the
cosmological constant, and $k$ is the sole remaining dimensionless coupling.
The action (\ref{3a1}) is invariant under the gauge invariance $A\rt  A + d\Lambda_A$,
$C \rt C+ d \Lambda_C$, up to a boundary term generated by $S_{CS}$; in fact,
(\ref{3a1}) is the most general two-derivative action with this property.
The field strength and potential forms, and their components, are related as follows,
\bea
\label{3a3}
A = A_M dx^M \hskip 0.65in & \hskip 0.7in & F = dA=  \half F_{MN} \, dx^M \wedge dx^N
\no \\
C =  \half C_{MN} dx^M \wedge dx^N &&
G= dC= {1 \over 6} G_{MNP} \, dx^M \wedge dx^N \wedge dx^P
\eea
Einstein's equations may be cast in the following form,
\bea
\label{3a4}
R_{MN} = - 2 F_{MP} F_N{}^P - G_{MPQ} G_N{}^{PQ}
+ g_{MN} \left ( {5 \over L^2} + {1 \over 4}  F_{PQ} F^{PQ} +{1 \over 6} G_{PQR} G^{PQR} \right )
\eea
Note that the contribution of the $F$-field is traceless in 4 dimensions while that of the $G$-field is
traceless in 6 dimensions, as we should expect. The Maxwell equations in the presence of the
2-form field $C$ are given by the Bianchi identity $dF=0$, together with the field equation,
\bea
\label{3a5}
\nabla _M F^{MN} = - { k \over 2 \, \sqrt{g}} \, \ep ^{MNPQRS} F_{PQ} \, \p_M C_{RS}
\eea
The 2-form field equations in the presence of the Maxwell field are given by the Bianchi identity $dG=0$,
together with the field equation,
\bea
\label{3a6}
\nabla _P G^{PMN} =  {k \over 4 \, \sqrt{g}} \ep ^{MNPQRS} F_{PQ} F_{RS}
\eea
In form notation, these equations become,
\bea
\label{3a7}
d*F - 2k F \wedge G & = & 0
\no \\
d *G + k F \wedge F & = & 0
\eea
Henceforth, we set $L=1$, and refer to the fields $F,G$ as the {\sl gauge fields} of the system,
and to (\ref{3a7}) as the {\sl gauge field equations}.

\subsection{Ansatz with translation and rotation symmetries}

The precise symmetry requirements will depend upon the physical questions addressed.
Throughout, we shall be interested in thermodynamic quantities (postponing problems
involving correlators to later work)
evaluated in the presence of a uniform magnetic $F_{34}=B$ field in the $x^3, x^4$ directions,
as well as a physical electric charge density $\rho$. Therefore, the natural symmetries to be imposed
will always include the following,\footnote{The field equations are also invariant under the
sign-reversal of $F$. We will not be led to impose this symmetry on our holographic solutions,
since we know that this discrete symmetry will be broken by a finite electric charge density.}
\begin{enumerate}
\itemsep -0.05in
\item Translations in $t, x^1,x^2, x^3, x^4$;
\item Rotations in the $x^1 x^2$-plane;
\item Rotations in the $x^3x^4$-plane.
\end{enumerate}
The Ansatz for the general metric invariant under the above symmetries is given as follows,
\bea
\label{3b1}
ds^2 ={dr^2\over U} -Udt^2 + e^{2V_1}\Big((dx^1)^2 +(dx^2)^2\Big )
+  e^{2V_2}\Big((dx^3)^2 +(dx^4)^2\Big)
\eea
This metric is expressed in a specific gauge choice for the holographic coordinate $r$
in which $g_{rr} g_{tt}=-1$. The functions $U, V_1, V_2$ depend only on $r$ by
translation invariance. Subject to the same symmetries, the gauge field Ans\"atze are given by,
\bea
\label{3b2}
F & = & E dr \wedge dt + B_1 dx^1 \wedge dx^2 + B_2 dx^3 \wedge dx^4
\no \\
G & = & \Big ( G_1 dr + G_2 dt \Big ) \wedge dx^1 \wedge dx^2 +
\Big ( G_3  dr + G_4 dt \Big ) \wedge dx^3 \wedge dx^4
\eea
Translation symmetry again restricts all coefficients to be independent of $x^0, x^1, x^2, x^3, x^4$.
To satisfy the Bianchi identities $dF=dG=0$ the coefficients  $B_1, B_2, G_2, G_4$ must be
independent of $r$. The remaining coefficients $E, G_1, G_3$ are generally functions of $r$.

\sm

Throughout, one of the magnetic fields will be kept non-zero, and used to generate the
RG flow from AdS$_6$ to a 2+1-dimensional IR boundary theory. The other magnetic field, which lives
in this 2+1-dimensional IR theory, will be set to zero for simplicity. We choose,
\bea
\label{3b3}
B_1=0  \hskip 1in B_2 = B
\eea
where $B \not= 0$. (Note that for $B_1=B_2$, exact solutions may be found which are closely related to
the dyonic AdS$_4$ Reissner-Nordstrom black brane \cite{D'Hoker:2009mm}.)

\subsection{General reduced gauge field equations}

It is straightforward to compute the reduced field
equations for the gauge fields, and we find,
\bea
\label{3c1}
M1 & \hskip 0.3in & 0 = \left ( E e^{2V_1+2V_2} \right ) ' + 2k BG_1
\no \\
M2 &  & 0 =  \left ( G_1 U   e^{2V_2-2V_1} \right ) ' +  2kB  E
\no \\
M3 &  & 0 =  \left ( G_3  U e^{2V_1 -2V_2}  \right )'
\eea
as well as the relation,
\bea
\label{3c2}
2kB G_2 = 0
\eea
Note that the Bianchi identities were solved already in the preceding subsection.
Since we are assuming throughout that $kB \not=0$ the field equations imply $G_2=0$.
Furthermore, equation $M3$  may be integrated once, to obtain,
\bea
\label{3c3}
G_3(r) = {c_3 \over U(r)} e^{-2V_1(r)+2V_2(r)}
\eea
where $c_3$ is an arbitrary integration constant.

\subsection{General reduced Einstein equations}

The reduced Einstein equations were computed using Maple.
The equation  for the $R_{rt}$ component is simply given by,
\bea
\label{3d1}
2G_1 G_2 e^{-4V_1} + G_3 G_4 e^{-4V_2}=0
\eea
Since we have already deduced the relation $G_2=0$ from the reduced gauge field
equations, we conclude that we must have $G_3 G_4=0$. With the help of (\ref{3c1}), this condition
becomes a relation between constants,
\bea
\label{3d2}
c_3 \, G_4=0
\eea
Eliminating $G_3(r)$ using equation (\ref{3c1}), and replacing the $R_{tt}$ equation by the combination
$(R_{rr}-R_{tt})/(4U)$, we find the following reduced Einstein equations,
\bea
\label{3d3}
E1 & \hskip 0.2in & 0 =
V_1'' +(V_1')^2 +V_2'' + (V_2')^2 + G_1^2 e^{-4V_1}
\no \\ && \hskip 0.2in
+ U^{-2} \left ( c_3^2 \, e^{-4V_1} + G_4^2 \, e^{-4V_2} \right )
\no \\
E2 && 0 =
U V_1 '' + U'V_1' + 2U V_1'^2  +2U V_1'  V_2'
-5 + \half E^2 - {1\over 2} B^2 e^{-4V_2}
\no \\ && \hskip 0.2in
+ U G_1^2 e^{-4V_1} - U^{-1} \left ( c_3^2 \, e^{-4V_1} -G_4^2 \, e^{-4V_2} \right )
\no \\
E3 & & 0 =
U V_2 '' + U'V_2' +  2U V_2'^2  +2U V_2'  V_1'
-5 + \half E^2 + {3\over 2} B^2 e^{-4V_2}
\no \\ && \hskip 0.2in
 - U G_1^2 e^{-4V_1} + U^{-1} \left ( c_3^2 \, e^{-4V_1} -G_4^2 \, e^{-4V_2} \right )
\no \\
E4 & & 0 =
U'' +2U'(V_1'+V_2') - 10 - 3 E^2 - B^2 e^{-4V_2}
\no \\ && \hskip 0.2in
-2 UG_1^2 e^{-4V_1} -2U^{-1} \left ( c_3^2 \, e^{-4V_1} + G_4^2 \, e^{-4V_2} \right )
\eea
The constraint equation is obtained as the combination $E2+E3-UE1$, and is given by
\bea
\label{3d4}
CON & \hskip 0.2in & 0 =
U'(V_1'+V_2') +U(V_1'^2+V_2'^2+4V_1' V_2') - 10  +E^2 +B^2 e^{-4V_2}
\no \\ && \hskip 0.2in
 - UG_1^2 e^{-4V_1} -U^{-1} \left ( c_3^2 \, e^{-4V_1} + G_4^2 \, e^{-4V_2} \right )
\hskip 1in
\eea
Using Maple, one can check that the derivative of the constraint equation $CON$
vanishes in view of the other Einstein and gauge field equations.

\subsection{Reduced equations for solutions with a smooth horizon}
\label{redsmooth}

We look for a smooth finite temperature horizon, at which $U$ has a single zero.
Other solutions will certainly exist, but it seems plausible that only solutions with smooth
horizons are of physical interest to us. From Einstein equation $E1$ in (\ref{3d3}) it is
clear that we need $c_3=G_4=0$ if we are to assume that $V_1$, $V_2$, and $G_1$ are
smooth functions at the horizon.  This condition, of course, automatically solves the Einstein
equation (\ref{3d2}). With this restriction, the gauge fields simplify as follows,
\bea
\label{3e1}
F & = & E(r) dr \wedge dt +  B dx^3 \wedge dx^4
\no \\
G & = &  G_1 (r) dr \wedge dx^1 \wedge dx^2
\eea
Equation $M3$ in (\ref{3c1}) is satisfied automatically, while equations $M1$ and $M2$ are unchanged;
we repeat them here for later convenience.
\bea
\label{3e2}
M1 & \hskip 0.3in & 0 = \left ( E e^{2V_1+2V_2} \right ) ' + 2k BG_1
\no \\
M2 &  & 0 =  \left ( G_1 U   e^{2V_2-2V_1} \right ) ' +  2kB \, E
\eea
With $c_3=G_4=0$, the reduced Einstein equations of (\ref{3d3}) simplify and become,
\bea
\label{3d3a}
E1 & \hskip 0.2in &
0 =  V_1'' +(V_1')^2 +V_2'' + (V_2')^2 + G_1^2 e^{-4V_1}
\no \\
E2 &&
0 =  U V_1 '' + U'V_1' + 2U V_1'^2  +2U V_1'  V_2'
-5 + \half E^2 - {1\over 2} B^2 e^{-4V_2}  + U G_1^2 e^{-4V_1}
\no \\
E3 & &
0 =  U V_2 '' + U'V_2' +  2U V_2'^2  +2U V_2'  V_1'
-5 + \half E^2 + {3\over 2} B^2 e^{-4V_2} - U G_1^2 e^{-4V_1}
\no \\
E4 & &
0 = U'' +2U'(V_1'+V_2') - 10 - 3 E^2 - B^2 e^{-4V_2} -2 UG_1^2 e^{-4V_1}
\eea
while the constraint is given by,
\bea
CON \hskip 0.2in
0 = U'(V_1'+V_2') +U(V_1'^2+V_2'^2+4V_1' V_2') - 10  +E^2 +B^2 e^{-4V_2} - UG_1^2 e^{-4V_1} \qquad
\eea

\subsection{Horizon Data}
\label{hordata}

For the purpose of carrying out numerical analysis,
we need to fix the remaining freedom in the choice of coordinates.
We put the horizon at $r=0$.  Using the freedom to rescale time~$t$
(while compensating by a rescaling of $r$),  as well as $x^{1,2}$ and $x^{3,4}$,
we shall normalize the components of the metric by setting,
\bea
U(0)=V_1(0)= V_2(0)=0   & \hskip 1in &     U'(0)=1
\eea
and use the following notation for the components of the gauge fields,
\bea
B=b \hskip 1in E(0) = q  \hskip 1in  G_1(0)=g
\eea
Solving $E2$, $E3$, $CON$, and $M2$ at the horizon produces expressions for the derivatives
$V_1'(0), V_2'(0)$, as well as a relation giving $g$ in terms of the other data,
\bea
g = -2k bq & \hskip 1in & V_1'(0) = 5-{1\over 2}q^2 +{1\over 2}b^2
\no \\
&& V_2'(0) = 5-{1\over 2}q^2 -{3\over 2}b^2
\eea
Thus, the free parameters, as specified at the horizon, are now $b$ and $q$.
We should find a unique solution for each choice of $(b,q)$, though the requirement
of global regularity will restrict the allowed range of these parameters.

\subsection{Asymptotic data and physical  quantities}
\label{asymdata}

For every regular solution, with specified values of $(b,q)$, the asymptotic behavior as $r \to \infty$ will
approach the geometry of the boundary of AdS$_6$. The $r \to \infty$ asymptotic behavior of the
components of the metric takes the familiar AdS$_6$ form,
\bea
{U(r) \over r^2} \to 1 \hskip 1in
{e^{2V_1(r)} \over r^2} \to v_1 \hskip 1in
{e^{2V_2(r)} \over r^2} \to v_2
\eea
The asymptotics of the gauge field strengths fix the physical charges of the
boundary theory,
\bea
E(r) \, r^4  \to q_\infty \hskip 1in
G_1(r) \, r^2  \to g_\infty
\eea
Physical quantities, namely the electric charge density $\rho$, the temperature $T$,
the entropy density $s$, the magnetic field $B_p$ and the 3-form charge density $G_p$ are
obtained as follows,
\bea
\rho = q_\infty \hskip 1in  T ={1\over 4\pi} \hskip 0.2in & \hskip 1in & B_p = {b\over v_2}
\no \\
s={1\over 4 v_1 v_2} ~ && G_p = {g_\infty \over v_1}
\label{phys}
\eea
In our definition of the entropy density we are absorbing Newton's constant:  $s = (G_6 S)/{\rm Vol}$.
We also comment on the form of $\rho$.  The electric charge of the system is given by
$\int \star F$, where the integral is over a constant time slice at the boundary.
Since $F = E dr\wedge dt$ we have $\star F \sim  v_1 v_2 r^4 E dx^1 \wedge dx^2 \wedge dx^3 \wedge dx^4$.   Dividing the integrand by $v_1 v_2$, to obtain the proper charge density, gives $q_\infty$.  Note that the temperature in these coordinates is fixed at the constant value $T=1/(4\pi)$. Since $T$ is dimensionful we can always scale coordinates to bring any nonzero temperature to this value.  Of course, this still allows the  physically relevant dimensionless temperature (denoted  below as $\hat{T}$) to vary.

\sm

As always, the only physically observable quantities in the conformal boundary theory are dimensionless
quantities. The above physical quantities all have non-trivial scaling dimensions under $ r \to \ell r$,
and $x^\mu \to x^\mu/\ell$, which are given by,
\bea
\rho \to \ell ^4 \rho \hskip 1in  T \to \ell T & \hskip 1in & B_p \to \ell^2 B_p
\no \\
s \to \ell^4 s && G_p \to \ell^3 G_p
\label{scaling}
\eea
Thus, only suitable dimensionless ratios can be physical observables. When $B_p \not= 0$,
as will be almost always the case here, we may use $B_b$ to obtain dimensionless physical
observables,
\bea
\label{3g5}
\hat T = {T\over \sqrt{B_p} } = { \sqrt{v_2} \over 4\pi \sqrt{b}}
& \hskip 1in &
\hat B = {B_p \over \sqrt{\rho} }  = {b\over v_2 \sqrt{q_\infty} }
\no \\
\hat s = {s \over B_p^2}= {v_2 \over 4 v_1 b^2} \hskip 0.19in
&&
\hat g = {G_p \over \sqrt{B_p^3}}= {g_\infty  \sqrt{v_2^3} \over  v_1 \sqrt{b^3} }
\eea
Given that the regular supergravity solutions are completely specified by the two parameters
$(b,q)$ of the near-horizon data (whose domain is governed by regularity restrictions),
together with the unique dimensionless coupling $k$, it follows that amongst the physical data
$\hat T, \hat B, \hat s$, and $\hat g$,
there must exist two dynamical relations, or {\sl equations of state}. Choosing $\hat T$ and $\hat B$
as independent parameters (as well as $k$), the dynamics of the system then determines,
\bea
\hat s & = & \hat s ( k, \hat T, \hat B )
\no \\
\hat g & = & \hat g ( k, \hat T, \hat B )
\eea
These functions will be obtained numerically by solving for $\hat T, \hat B, \hat s$, and $\hat g$
in terms of the near-horizon parameters $(b,q)$.

%\newpage

%%%%%%%%%%%%%%%%%%%%%%%%%%%%%%%%%%%%%%%%%%%
%%%%%%%%%%%%%%%%%%%%%%%%%%%%%%%%%%%%%%%%%%%
\section{Purely Electric and Purely Magnetic solutions}
\setcounter{equation}{0}
\label{four}
%%%%%%%%%%%%%%%%%%%%%%%%%%%%%%%%%%%%%%%%%%%
%%%%%%%%%%%%%%%%%%%%%%%%%%%%%%%%%%%%%%%%%%%

In this section, we shall briefly review the AdS$_6$ Reissner-Nordstrom (RN) solution
for which no magnetic field is present, $B=0$. For our purposes the RN solution  is perhaps better referred
to as the {\sl purely electric solution}. We shall then exhibit an exact near-horizon solution
with $B \not= 0$ but zero electric charge, whose metric is that of AdS$_4 \times \bR^2$.
The {\sl purely magnetic solution} is the unique interpolating solution between this
near-horizon solution and asymptotic AdS$_6$. We shall show, numerically, that the purely
magnetic solution exists and we shall characterize its asymptotics.

\subsection{The purely electric solution}
\label{foura}

In the standard coordinates, the RN solution is given by,
\bea
\label{4a1}
U = r^2+{q^2 \over 6 r^6}-{M\over r^3} \hskip 0.8in
e^{2V_1}=e^{2V_2}=r^2 \hskip 0.8in
E = {q \over r^4}
\eea
In the extremal limit, the location of the horizon $r_+$ obeys $U(r_+)=U'(r_+)=0$, which
requires the following relations between $q,M$ and $r_+$,
\bea
\label{4a2}
q= \sqrt{10} \, r_+^4 \hskip 1in M = { 8 \over 3} \, r_+^5
\eea
We see that the extremal entropy density $s$ is proportional to  the charge density, $s=r_+^4 \propto q$,
and does not vanish as $T=0$.

\sm

In the coordinates described in section  \ref{hordata} the RN solution is
given by,
\bea
\label{4a3}
E = {q \over (Cr +1)^4} \hskip 1in  e^{2V_1} = e^{2V_2} = (Cr+1)^2
\eea
and $U$ takes the form,
\bea
\label{4a4}
U= { u_0 \over (Cr+1)^3} + {1 \over C^2} (Cr +1)^2 + { q^2 \over 2 C^2} { 1 \over (Cr+1)^6}
\eea
with
\bea
\label{4a5}
C = {10-q^2 \over 2} \hskip 1in u_0 = - { 2+q^2 \over 2 C^2}
\eea
The extremal limit is obtained at $q^2 =10$. The near-horizon geometry of the purely electric
solution is AdS$_2 \times \bR^4$.

\subsection{The AdS$_4 \times \bR^2$ near-horizon solution}
\label{fourb}

Turning on the magnetic field modifies the near-horizon geometry. Simplest is the purely magnetic
case for which the electric charge vanishes. Here, we can obtain a fairly simple picture that will
be relevant for interpreting the numerics.  Setting $G_1 = E=0$, a simple solution which is regular
at the horizon is then given by,
\bea
\label{4b1}
U = U_0 r^2 \hskip 0.8in
e^{2V_1}=r^2 \hskip 0.8in
e^{2V_2}=1
\eea
where the constants are given by,
\bea
\label{4b2}
b=\sqrt{10\over 3} \hskip 0.8in  U_0  = {20 \over 9}
\eea
This geometry is recognized as AdS$_4 \times \bR^2$, and is the $D=6$ analog of the AdS$_3 \times \bR^2$ solutions studied in \cite{D'Hoker:2009mm,D'Hoker:2010hr}.  Magnetic brane solutions have been studied further in   \cite{Almuhairi:2010rb,Almuhairi:2011ws,Donos:2011qt,Donos:2011pn,Almheiri:2011cb},
including supersymmetric examples.  One lesson is that when these solutions are embedded into a higher dimensional supergravity theory they become subject to various potential instabilities, and the same could be expected of the AdS$_4\times \bR^2$ solution discussed here.

\subsubsection{Perturbing around the AdS$_4 \times \bR^2$ near-horizon solution}
\label{fourc}

To construct solutions interpolating between AdS$_4 \times \bR^2$ and AdS$_6$ we proceed by including small perturbations around  AdS$_4 \times \bR^2$ and using these to seed the flow towards AdS$_6$.
Since the  AdS$_4 \times \bR^2$ near-horizon solution has $E=G_1=0$, the linearized
Einstein and gauge field equations around this solution decouple from one another.

\sm

We begin by parameterizing the small fluctuations fields for the metric as follows,
\bea
\label{4c1}
U(r) & = & U_0 r^2 (1 + 2   u(r) )
\no \\
V_1(r) & = & \half \ln (U_0 r^2) + v_1(r)
\no \\
V_2(r) & = &  v_2(r)
\eea
where the fields $u,v_1, v_2$ are treated as first order perturbations. One obtains a closed
equation for $v_2$ from $E3$, then an equation for $v_1$ from $E1$, and finally an equation for
$u$ from $CON$. Putting all together, we find the following system for the Einstein equations,
\bea
\label{4c2}
r^2 v_2'' + 4 r v_2' - 9 v_2 & = & 0
\no \\
r^2 v_1'' + 2 r v_1' + r^2 v_2'' & = & 0
\no \\
2ru' + 6 u + 4 r v_1' + 6 r v_2' - 6 v_2 & = & 0
\eea
The equations are invariant under dilations in $r$, as were the original non-linear
equations of (\ref{3d3}), and the solutions to the linear equations
are linear combinations of powers of $r$. For $v_2(r) \sim r^\sigma$, we must have
\bea
\label{4c3}
\sigma^2 +3 \sigma -9=0 \hskip 1in \sigma = {3 \over 2} ( -1 + \sqrt{5} )
\eea
The positive root for $\sigma$ has been retained here because we are after a regular
perturbation near $r=0$, requiring $v_2$ to vanish there. The regular solutions for
the other fields are now easily found to be,
\bea
\label{4c4}
u(r)=v_1(r) = -{ \sigma -1 \over \sigma +1} \, (c r)^\sigma \hskip 1in v_2(r) = (c r)^\sigma
\eea
Here, $c$ is an arbitrary integration constant, or scale parameter, which may of course
be absorbed into the definition of $r$. Equality of $u$ and $v_1$ guarantees
2+1-dimensional Poincar\'e invariances, as expected.

\sm

Since the gauge fields of the AdS$_4 \times \bR^2$ near-horizon solution vanish,
the fields $E, G_1$ in the gauge field equations may be viewed themselves as
parametrizing the first order fluctuations, and thus obey the linearized equations,
\bea
\label{4c5}
\left ( U_0 r^2 E \right ) ' +  kb G_1 & = & 0
\no \\
G_1' + 2 kb E & = & 0
\eea
where the constants $b$ and $U_0$ are given by (\ref{4b2}).
Eliminating $G_1$, we find the following equation $(r^2 E)'' - 6 k^2 E=0$ for $E$.
This is, of course,  again scale invariant, so that its solutions are linear combinations
of powers of $r$. The full solution is found to be given by,
\bea
\label{4c6}
E(r) = r ^\kappa
\hskip 1in
G_1(r) = - {  (\kappa +2)b \over 3 k} \, r^{\kappa +1}
\eea
where the exponent $\kappa$ must satisfy,
\bea
\label{4c7}
(\kappa +2)(\kappa +1) - 6 k^2 =0
\hskip 1in \kappa = -{ 3 \over 2} + \half \sqrt{1 + 24 k^2}
\eea
The positive square root chosen above is dictated by the regularity at the horizon of the gauge
potential $A$, which is defined by $E=A'$ together with its vanishing at the horizon. Note that we then
have $\kappa \geq -1$ for all $k$, with equality only for $k=0$. The electric field itself is regular at
the horizon provided $\kappa \geq 0$, namely when
\bea
k \geq k_c= {1 \over \sqrt{3}}
\eea
This condition singles out the value $k_c$ for the coupling $k$
(which will turn out to be a {\sl critical value})
as a rough analogue of the special value $k=1$ for the AdS$_5$ case \cite{D'Hoker:2010ij}.

\subsection{The purely magnetic brane solution}

We have not succeeded in solving analytically the reduced Einstein equations for the
purely magnetic brane (the gauge field equations of (\ref{3e2}) are satisfied automatically for
zero charge when $E=G_1=0$). Numerical analysis is, however, straightforward and shows
that the first order perturbation of the metric around the AdS$_4 \times \bR^2$ near-horizon solution,
computed  in section \ref{fourc}, extends into a full fledged solution of the non-linear equations
which interpolates to asymptotically AdS$_6$. Given the above normalizations at the horizon,
the AdS$_6$ data result as follows,
\bea
{ U(r) \over r^2}  \to 1 \hskip 0.8in
{ e^{2V_1(r)} \over r^2}  \to 1 \hskip 0.8in
{ e^{V_2(r)} \over r^2} =  C_v
\eea
and $C_v$ is obtained numerically, and given by $C_v = 1.221905$.

%\begin{figure}[htb]
%\begin{centering}
%\includegraphics[scale=0.7]{AsymB_fig3}
%\caption{Data in red dots, and fit $-\ln \hat s = -\ln s_0 - \ep \ln %\hat T$ in solid blue line,
%in the regime of small  $T^2/ B$ over a range $5.7 < \hat B < 2.3 %\times 10^9$. The best fit is $ s_0=1.17$,
%and $\ep =2.0084$. In the figure, $S,T$ stand for $\hat s, \hat T$ %respectively.}
%\label{flow3}
%\end{centering}
%\end{figure}
%

%\subsection{The entropy density for small $T^2/B$}
%
%The small $T^2/B$ behavior of the entropy density is dominated by %the purely magnetic
%brane whose horizon, we have established, is $AdS_4 \times \bR^2$. %Scaling arguments
%then suggest that the entropy density must behave like $T^2$ for %small $T$. On dimensional
%grounds, we then obtain the following relation for the entropy %density, valid for small $T^2/B$,
%\bea
%\hat s =  s_0 \hat T^2 \qquad\quad \hbox{or} \qquad\quad  s = s_0 B %T^2
%\eea
%with the constant $s_0$ governed by the purely magnetic brane %solution. This relation, obtained
%on the grounds of general arguments, may be checked by numerical %analysis.
%One may proceed, when working numerically, by keeping a small charge %density,
%and then establishing independence of the value of this density. %This gives the regime
%of small $T^2/B$, but with proper ``regularization". The data and fit are given in figure \ref{flow3}.
%The $AdS_4$ relation holds over 7 e-folds, and over a wide range of values of $\hat B$.
%The constancy over this wide range allows us to determine the constant $s_0$ as follows,
%\bea
%s_0 \sim 1.17
%\eea

\subsection{Entropy density of purely magnetic brane for small $T^2/B$}

In the purely magnetic case an expression for the low temperature entropy density may
be obtained analytically, as we now discuss.  This provides a useful check on the numerics.

\sm

The purely magnetic brane solution takes the form,
\bea  ds^2 &=& {dr^2 \over U}- Udt^2 +e^{2V_1} \Big((dx^1)^2+(dx^2)^2\Big)+e^{2V_2}\Big((dx^3)^2+ (dx^4)^2\Big) \cr
F&=& B dx^3 \wedge dx^4
\eea
At zero temperature we have Lorentz invariance in $(t,x^1,x^2)$, and this  fixes $e^{2V_1}=U$.
The zero temperature solution has near-horizon geometry AdS$_4 \times \bR^2$, so that,
as $r \to 0$,
\bea
U = U_0 r^2
\hskip 1in
e^{2V_2} = \sqrt{3\over 10} B
\eea
with $U_0=20/9$, as was given in (\ref{4b2}).

\sm

At small but finite temperature the near-horizon AdS$_4$ is replaced by the near-horizon AdS$_4$
Schwarzschild solution, given by,
\bea
U =  U_0\left(r^2 - {r_+^3\over r} \right)
\hskip 1in
e^{2V_1}  =  U_0 r^2
\hskip 1in
e^{2V_2}  =  \sqrt{3\over 10}B
\eea
This is valid provided the horizon lies well within the AdS$_4$ region, namely for $r_+ \ll 1$.
This translates into the condition $T^2 / B \ll 1$.
The entropy density and temperature are
\bea
 G_6 S = {A_H \over 4} = {1\over 4} \sqrt{3\over 10} U_0 B r_+^2 V_4
 \hskip 1in
 T= {U'(r_+)\over 4\pi} = {3U_0 r_+ \over 4\pi}
 \eea
where $V_4$ is the coordinate volume.  Eliminating $r_+$ between these expressions,
we compute the entropy density $s$ as a function of temperature,
\bea
\label{pureBs}
s = {G_6 S \over V_4} = \sqrt{3\over 10} {\pi^2 \over 5} BT^2  \, \approx \, 1.08 \, B \, T^2
\eea
The $T^2$ dependence is characteristic of a $2+1$ dimensional CFT; this CFT is associated
to the near-horizon AdS$_4$ region.

%%%%%%%%%%%%%%%%%%%%%%%%%%%%%%%%%%%%%%%%%%%
%%%%%%%%%%%%%%%%%%%%%%%%%%%%%%%%%%%%%%%%%%%
\section{Lifshitz solutions for critical $k$}
\setcounter{equation}{0}
\label{five}
%%%%%%%%%%%%%%%%%%%%%%%%%%%%%%%%%%%%%%%%%%%
%%%%%%%%%%%%%%%%%%%%%%%%%%%%%%%%%%%%%%%%%%%

The near-horizon Reissner-Nordstrom black brane geometry for $k<k_c$,
and the near-horizon AdS$_4\times \bR^2$ for $k>k_c$ are separated by
a Lifshitz near-horizon geometry for $k=k_c$. The fields for the corresponding
near-horizon solution will be computed analytically in this section. Linearized
fluctuations around this solution smoothly connect onto solutions to the full fledged
non-linear reduced field equations which interpolate to AdS$_6$, as will be shown
in this section, using the results of Appendix A.

\subsection{Near horizon Lifshitz solution}
\label{lifshitz}

In this sub-section we look for scale invariant solutions that can play the role of near-horizon
geometries in the presence of nonzero electric charge and magnetic field. We start from the Ansatz
arrived at in section \ref{redsmooth} for the metric,
\bea
\label{metricc}
ds^2 ={dr^2\over U} -Udt^2 + e^{2V_1}\Big((dx^1)^2 +(dx^2)^2\Big )
+  e^{2V_2}\Big((dx^3)^2 +(dx^4)^2\Big)
\eea
and for the gauge fields,
\bea
F & = & E(r) dr \wedge dt +  b_0 \, dx^3 \wedge dx^4
\no \\
G & = &  G_1 (r) dr \wedge dx^1 \wedge dx^2
\eea
We look for a solution which is invariant under scale transformations of the form,
\bea
\label{scaleeqs}
r\rt \lambda r
\hskip 0.8in
t \rt {1\over \lambda}t
\hskip 0.8in
x^{1,2} \rt {1\over \lambda^{\alpha}} x^{1,2}
\hskip 0.8in
x^{3,4} \rt {1\over \lambda^{\beta}} x^{3,4}
\eea
for some real constants $\alpha$ and $\beta$. Scale invariance requires,
\bea
U = u_0 r^2
\hskip 0.6in
e^{2V_1} = r^{2\alpha}
\hskip 0.6in
e^{2V_2} = r^{2\beta}
\hskip 0.6in
E=q_0
\hskip 0.6in
G_1 = g_1 r^{2\alpha -1}
\eea
where we used the freedom to rescale $x^i$ to fix the prefactors of
$e^{2V_{1,2}}$.   The metric is of the Lifshitz type \cite{Kachru:2008yh}.
We now substitute these expressions into the reduced field equations written in
section \ref{redsmooth}.

Assuming $b_0\neq 0$, equation $E2$ implies $\beta=0$, since all terms in $E2$ are
constant except for the final term, which behaves as $b_0^2 r^{-4\beta}$.
Solving $M1$ and $M2$ requires,
\bea
u_0= {2k^2 b_0^2 \over \alpha}
\hskip 1in
g_1 = -{q_0 \alpha \over kb_0}
\eea
Solving $E1$ gives,
\bea
q_0^2 = \left( 1-\alpha \over\alpha\right) k^2 b_0^2
\eea
We then find,
\bea
E2-E3 = 2b_0^2 (3k^2 -1)
\eea
which, assuming as always $b_0\neq 0$,  forces,
\bea
k = k_c = {1\over \sqrt{3}}
\eea
The remaining equations reduce to the following quadratic equation for $\alpha$,
\bea
\left ( \alpha+{1\over 2} \right )^2 = {15 \over 2b_0^2}\alpha
\eea
Its solutions may be parameterized as follows,
\bea
\label{qbvals}
q_0^2 =  {10 z(z-1) \over (z+2)^2}
\hskip 1in
b_0^2 ={ 30z \over (z+2)^2}
\eea
where we have defined,
\bea  z ={1\over \alpha}
\eea
From (\ref{scaleeqs}) we see that $z$ is the dynamical critical exponent,
since under scaling $t \sim (x^{1,2})^z$.

\sm

We have therefore found a one parameter family of solutions Lif$_4 \times \bR^2$
parameterized by the dynamical exponent $z$.   From (\ref{qbvals}) it is clear that we should
restrict to $z\geq 1$ in order that $q_0$ and $b_0$  be real.   All values $z\geq 1$ are allowed.
For $z=1$ we recover the purely magnetic AdS$_4 \times \bR^2$ solution, with
$(b_0,q_0) = (\sqrt{10/3},0)$.    As $z\rt \infty$ we go over to the purely electric
AdS$_2 \times \bR^4$ solution with $(b_0,q_0) =(0, \sqrt{10})$.
For $k\neq k_c$, only the purely magnetic/electric solutions exist.

\subsection{Interpolating solution}

We now look for solutions that interpolate between these $k=k_c$ near-horizon solutions
and an asymptotic AdS$_6$.  Here we proceed numerically.  We first solve for linearized
fluctuations around Lif$_4\times \bR^2$, identify those that are nonsingular in the near-horizon
region, and then use these as seeds for the numerical integration out to large $r$.
The details are described in Appendix \ref{AppA}.  The upshot is that we find smooth
asymptotically AdS$_6$ solutions for the full family of near-horizon solutions described above.

%\newpage

%%%%%%%%%%%%%%%%%%%%%%%%%%%%%%%%%%%%%%%%%%%
%%%%%%%%%%%%%%%%%%%%%%%%%%%%%%%%%%%%%%%%%%%
\section{Flows towards low temperature}
\setcounter{equation}{0}
\label{six}
%%%%%%%%%%%%%%%%%%%%%%%%%%%%%%%%%%%%%%%%%%%
%%%%%%%%%%%%%%%%%%%%%%%%%%%%%%%%%%%%%%%%%%%

We now study the  thermodynamics of solutions with nonzero charge and magnetic field.
In the high temperature limit the behavior is universal, and is governed by the AdS$_6$
Schwarzschild black brane.  The entropy density behaves as $s \propto T^4$, as
appropriate for a 4+1 dimensional CFT. The low temperature behavior is much more
interesting, as it is sensitive to the charge density, magnetic field, and Chern-Simons
coupling $k$.
We have already identified $k_c = 1/\sqrt{3}$ as the critical value of $k$ for which a
near-horizon Lifshitz geometry arises, and so we can expect a qualitative change in behavior
depending on the size of $k$ relative to $k_c$, as was summarized in the Introduction.
Here we supply more of the details.

\sm

The discussion in sections \ref{hordata} and \ref{asymdata} provides the basis for our
numerical evaluation of the entropy density.  As described there, after fixing coordinates
the near-horizon data is parameterized by specifying the pair $(b,q)$, representing the
near-horizon electric and magnetic fields.
Numerical integration outwards in $r$ for each choice of $(b,q)$  (and for a given $k$)
yields either a smooth asymptotically AdS$_6$ solution or a singularity.
We discard the singular solutions, which requires that we restrict attention to a finite
region in the $(b,q)$ plane.  For each $(b,q)$ within the allowed region we compute the
dimensionless quantities
$(\hat{T}, \hat{s},\hat{B}, \hat{g})$ from the asymptotic data at the AdS$_6$ boundary.
Our main interest is in understanding the behavior of $\hat{s}(\hat{T})$ at fixed $\hat{B}$.
As we have noted, this depends crucially on the size of $k$ relative to $k_c$, and we discuss
the three cases in turn.

\subsection{Low temperature thermodynamics: $k>k_c$}

\begin{figure}[htb]
\begin{centering}
\includegraphics[scale=1.0]{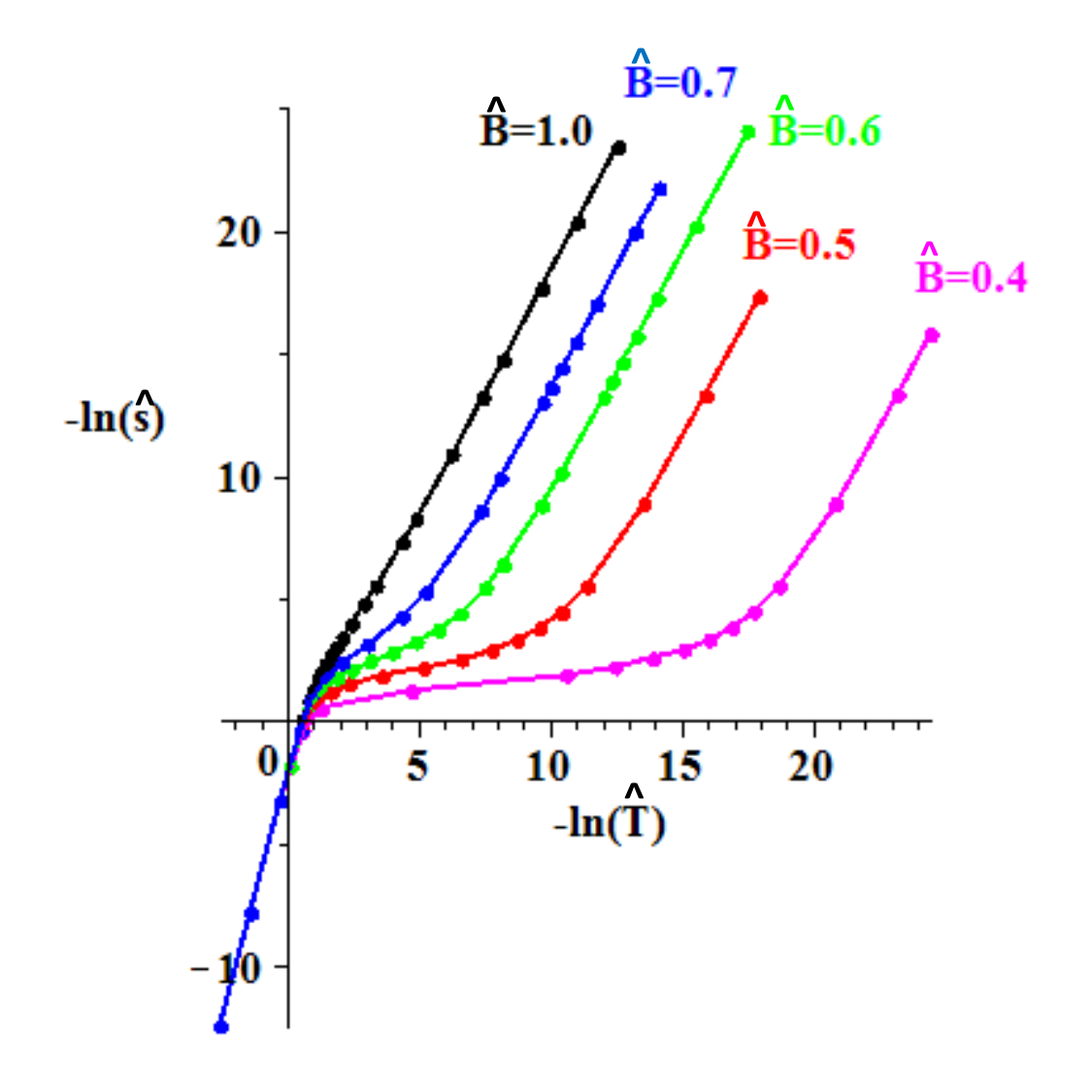}
\caption{Entropy density versus temperature at $k=0.7$ for selected values of $\hat{B}$. }
\label{figfive}
\end{centering}
\end{figure}

At vanishing charge density, corresponding to $\hat{B} \rt \infty$, the low temperature
entropy density was found analytically in (\ref{pureBs}).  In terms of dimensionless
quantities the result is,
\bea
\hat{s} = \sqrt{3\over 10}{\pi^2 \over 5}\hat{T}^2
\hskip 1in {\rm as}
\quad   \hat{T} \rt 0\quad {\rm at}\quad \hat{B} = \infty
\eea
More generally, at finite $\hat{B}$ the numerics show the low temperature behavior, as $\hat T \rt 0$,
\bea
\hat{s} = A(k,\hat{B}) \, \hat{T^2}
 \eea
Typical examples are shown in  figure \ref{figfive}.
\sm

The $\hat{T}^2$ dependence of the entropy density is associated with a near-horizon
AdS$_4\times \bR^2$ geometry, in particular the one discussed in section \ref{fourb}.
We note that the electric field vanishes in the near-horizon region, $E=0$, which makes
it clear that the electric flux measured at the AdS$_6$ boundary is sourced by fields living
outside the near-horizon region.   As emphasized in the Introduction, it is the modification of Gauss' law brought about by the Chern-Simons term that allows the combined magnetic field and three form field strength to act as a charge density.  For $k>k_c$, we see that all of the charge density measured at the AdS$_6$ boundary is generated by this mechanism.

\sm

Another way to  exhibit the appearance of AdS$_4\times \bR^2$ is to plot the flow in $(b,q)$
space as we lower the temperature, as shown in figure \ref{fig2}.  Lowering the temperature at fixed
$\hat{B}$, we find that we are driven to the point $(b= \sqrt{10/3}, q=0)$.
This point  represents the pure magnetic brane solution; of course, we never reach
precisely this point, since we are keeping $\hat{B}$ fixed at a finite value,  while the magnetic
brane has $\hat{B}=\infty$.

\sm

\begin{figure}[htb]
\centering
\begin{minipage}[c]{0.5\linewidth}
\centering \includegraphics[width=3.4in]{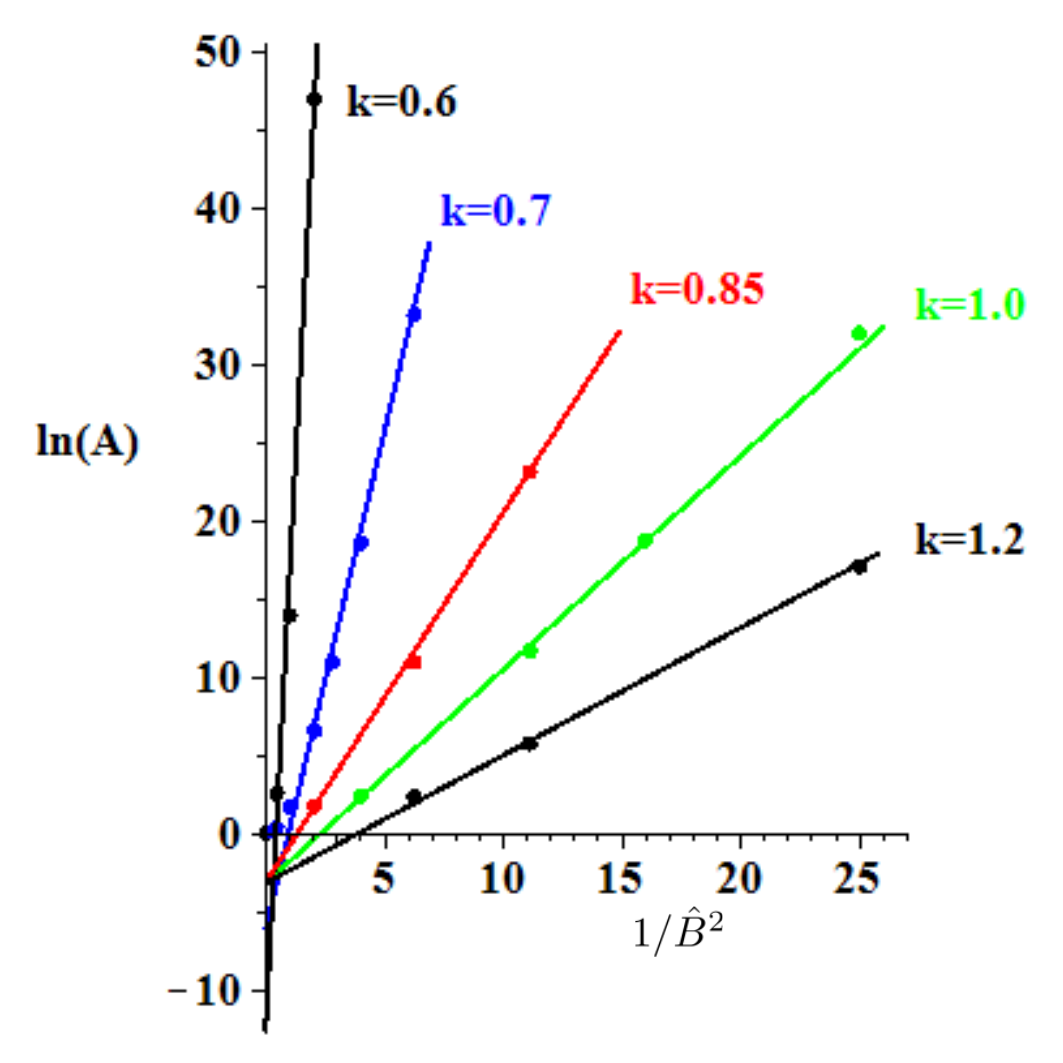}
\end{minipage}%
\begin{minipage}[c]{0.5\linewidth}
\centering \includegraphics[width=2.6in]{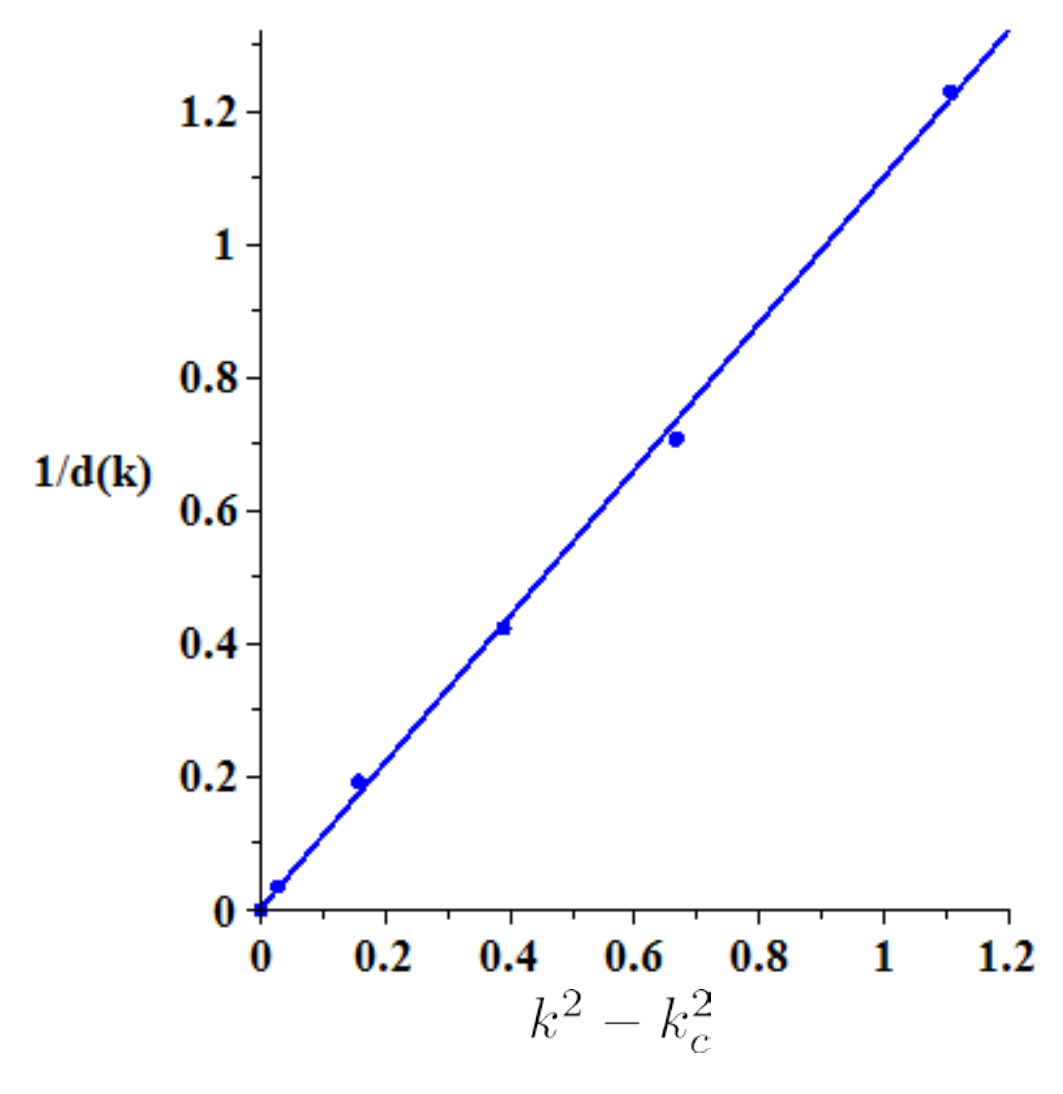}
\end{minipage}
\caption{Data illustrating the divergence of $A(k,
\hat{B})$ as $\hat{B}\rt 0$, in accord with (\ref{Adiv}). }
\label{dvsk}
\end{figure}

Next we consider decreasing $\hat{B}$.  Strictly at $\hat{B}=0$, we know that we should
recover the pure electric RN solution and its nonzero ground state entropy density,
and so it is clear that $A(k,\hat{B})$ must be singular at $\hat{B}=0$.
The numerics show the following behavior as $\hat B \rt 0$,
\bea A(k,\Bh)\sim   c(k) \exp \left \{ {d(k)\over   \Bh^2} \right \}
\label{Adiv}
\eea
for some $k$-dependent real positive constants $c$ and $d$; see figure \ref{dvsk}.
On general grounds we expect that this singularity structure should be amenable to
analytical understanding, but this is lacking at present.

\subsection{Low temperature thermodynamics: $k<k_c$}

For sufficiently small $k$ the effect of the  Chern-Simons term is too weak to allow for all the
charge density to be carried by flux.  Some fraction of the flux is instead hidden behind an
event horizon, and associated with this horizon is a nonzero ground state entropy density.
Indeed, for $k<k_c$ there appears to be such an entropy for any finite $\hat{B}$.
Representative numerical results are shown in figure \ref{fig2}.  Note that increasing
$\hat{B}$ suppresses the value of the extremal entropy.

\sm

The near-horizon geometry obtained at $\hat{T}=0$ in this regime is AdS$_2 \times \bR^4$.
This can also be seen by studying the flow in $(b,q)$ space. As $\hat{T}$ is lowered at fixed
$\hat{B}$, we flow towards $(b=0, q= \sqrt{10})$, which is the pure electric fixed point.
This is illustrated in Figure \ref{fig2}.

\sm

Increasing $k$ at fixed $\hat{B}$ again reveals $k_c$ to represent  a critical value. In particular,
the extremal entropy tends toward zero as $k$ approaches $k_c$ from below; this behavior
can be seen in figure \ref{fig2}.

\subsection{Low temperature thermodynamics: $k=k_c$}

At the critical value of $k$, taking the temperature to zero reveals the appearance of the
near-horizon Lif$_4 \times \bR^2$ geometry discussed in section \ref{five}.
Examining the flow in $(b,q)$ space as in figure \ref{fig2} shows that, rather than flowing to
either the pure electric or pure magnetic fixed point, the IR flow is towards a point on a critical curve.
To explain this we first note that according to (\ref{qbvals}) the parameters of the
$\Th=0$  Lif$_4 \times \bR^2$ solution obey,
\bea
3(q_0^2+b_0^2)^2 -10(b_0^2+3q_0^2)=0
\label{critcur}
\eea
This same curve governs the $\Th \rt 0$  behavior of possible values of $(b,q)$, up to the fact that there is a finite renormalization relating $(b,q)$ to $(b_0,q_0)$.   This renormalization comes about from the fact that $(b,q)$ are defined at the horizon of a finite temperature solution in a coordinate system (defined in section \ref{hordata}) that does not agree, in the zero temperature limit,  with that used to write the zero temperature solution in which $(b_0,q_0)$ appear.  The coordinate transformation that relates the two solution induces a change in the parameters.      However, as the temperature is taken to zero, the two coordinate systems differ only in a shrinking region near the horizon and, as long as one stays outside this region, the parameters $(b_0,q_0)$ can be read off from the finite temperature solution, and the critical curve (\ref{critcur}) is reproduced from the numerics. This procedure is explained in more detail Appendix \ref{AppB}.

\sm

\begin{figure}[htb]
\begin{centering}
\includegraphics[scale=0.8]{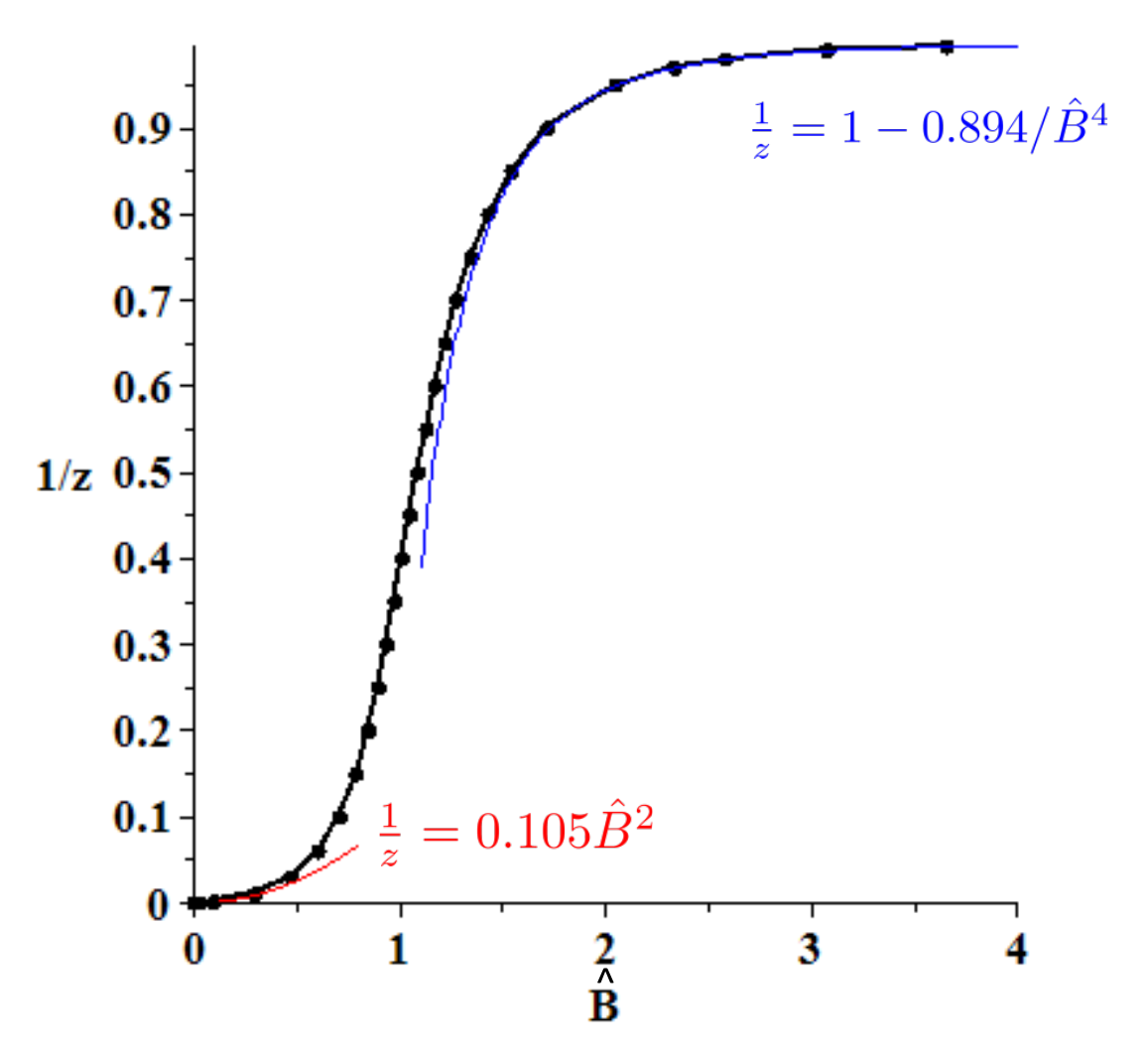}
\caption{Dynamical critical exponent $z=1/\a$ as a function of $\hat B$ for $k=k_c$.}
\label{critexp}
\end{centering}
\end{figure}

The critical  curve can be parameterized by the dynamical exponent
$z$, according to (\ref{qbvals}).
The critical curve connects the electric fixed point at $b=0$ and the magnetic fixed point at $q=0$.
The full curve is traversed as $\hat{B}$ ranges from $0$ to $\infty$.   There is thus a relation
between the dynamical exponent $z$ and the magnetic field,  $z=z(\hat{B})$.
However, the precise form of this relation depends on the full interpolating solution
connecting the near-horizon region to AdS$_6$.  This is because $\hat{B}$ is defined in
terms of asymptotic data, while $(b,q)$ are near-horizon data.
Numerically, we find that $z(\hat{B})$ takes the form shown in figure \ref{critexp}.

\sm

The low temperature entropy takes a form dictated by the near-horizon Lifshitz symmetry,
namely, as $\hat T \rt 0$ we have,
\bea
\hat{s} = f(\hat{B}) \, \hat{T}^{2/z} \hskip 1in z = z (\hat B)
\eea
We have not attempted to characterize the prefactor $f(\hat{B})$, but it presumably depends on the full interpolating solution.
\begin{figure}[htb]
\begin{centering}
\includegraphics[scale=0.8]{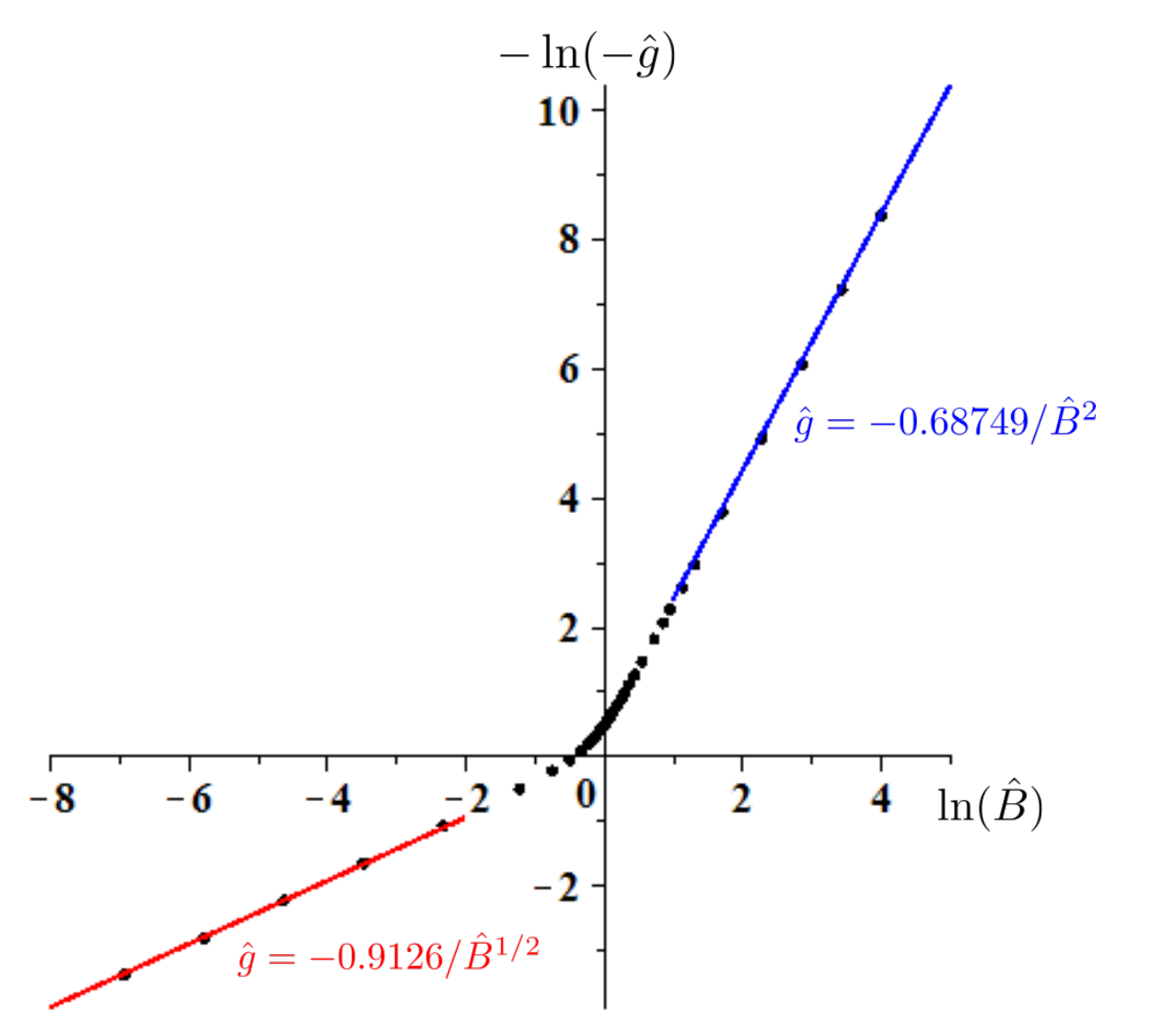}
\caption{Plot of the three-form charge $\hat{g}$ as a function of $\hat{B}$ at $T=0$ and $k=k_c$.}
\label{flow1}
\end{centering}
\end{figure}

These solutions also carry a nonzero value for the three-form charge $\hat{g}$.
Its behavior is shown in figure \ref{flow1}. The power law dependence for large and small $\hat{B}$ can be understood as follows.  First consider the $\hat{B}\rt 0$ behavior.  At $B=0$ we have the purely electric RN solution, which has $G=0$.   We can use perturbation theory around the purely electric solution to add in a small $B$.  Assuming a finite radius of convergence, we expect that the leading behavior of $G_p$ is linear in $B_p$, where we now refer to the physical quantities defined in (\ref{phys}).  The $\rho$ dependence is then fixed by the scale transformations defined in (\ref{scaling}), which give $G_p  \sim \rho^{1/4}B_p$.  Converting to hatted quantities, this is $\hat{g} \sim 1/\hat{B}^{1/2}$, which agrees with the behavior in figure  \ref{flow1}.        The same type of argument holds for $\hat{B}\rt \infty$, now involving perturbation theory around the purely magnetic solution.  The fact that the numerics agree with these arguments based on perturbation theory can be taken as evidence that perturbation theory is convergent.

%%%%%%%%%%%%%%%%%%%%%%%%%%%%%%%%%%%%%%%%%%%
%%%%%%%%%%%%%%%%%%%%%%%%%%%%%%%%%%%%%%%%%%%
\section{Quantum critical behavior}
\setcounter{equation}{0}
\label{seven}
%%%%%%%%%%%%%%%%%%%%%%%%%%%%%%%%%%%%%%%%%%%
%%%%%%%%%%%%%%%%%%%%%%%%%%%%%%%%%%%%%%%%%%%

Our analysis has revealed two interesting examples of quantum critical behavior in this system,
as manifested by the low temperature thermodynamics.    In particular, we focus on non-analytic
behavior in terms of the control parameters $k$ and $\hat{B}$, and the temperature $\hat{T}$.

\sm

In the regime $k>k_c$ we found nonanalytic behavior at $\hat{B}=0$.  This is associated with the fact that an infinitesimally small magnetic field is enough to remove the ground state entropy density of the pure electric RN black brane.    Quantitatively, we saw nonanalytic behavior in the coefficient of the $\hat{T}^2$ term in the entropy density as $\hat{B}\rt 0$.  As noted already, we lack a good analytical understanding of this critical behavior.

\sm

A second example of quantum criticality was found at $k=k_c$, and governed by the appearance of a near-horizon Lifshitz geometry.  We explored the approach to this critical point from several directions, either by decreasing $\hat{T}$ at fixed $\hat{B}$ and $k$, and by letting $k$ approach $k_c$ at fixed $\hat{B}$ and (low) $\hat{T}$.  More generally, there will exist a full scaling form for the entropy density in the vicinity of the critical point
\bea
\hat{s} = F(k-k_c,\hat{B},\hat{T})
\eea
In the analogous AdS$_5$ critical point the full scaling function was computed
analytically in \cite{D'Hoker:2010ij}, and it would of course be interesting to achieve that for
the present case as well, but we leave that for the future.

\sm

The two control parameters $k$ and $\hat{B}$ are on somewhat different footings in
the (nominal)  dual CFT. $\hat{B}$ is built out of the  magnetic field $B$ and charge
density $\rho$.  Changing $\rho$ corresponds to considering a different state of the CFT,
while changing $B$ is accomplished by varying an external field.  On the other hand, we
expect that to change $k$ we need to change the field content of the CFT.   Chern-Simons
terms in the bulk are related to anomalies in the dual CFT, and these are determined by the
field content.   It is therefore unclear whether we could ever realize the precise value $k_c$,
or whether it is physically sensible to think about tuning $k$.   On the other hand, the
familiar ``fanlike" structure governing finite temperature behavior near a quantum critical point
shows that it is not necessary to sit precisely at the critical point in order to detect its influence.

\sm

The critical D=2+1 theories studied here violate parity due to the three-form field strength
$G= G_1 dr \wedge dx^1 \wedge dx^2$.  In the boundary CFT, this corresponds to an
expectation value of some two-form operator, $\langle {\cal O}_{12} \rangle = -\langle {\cal O}_{21} \rangle$.   Since we haven't proposed a specific CFT dual, we of course cannot say what this operator is.
But given such a dual CFT, we expect that this two-form operator will play an important role in
driving the quantum phase transition.

\vskip 0.35in

\noindent
{\Large \bf Acknowledgments}

\vskip 0.15in

We thank Eric Perlmutter for discussions.
Eric D'Hoker wishes to thank the Laboratoire de Physique Th\'eorique del'Ecole Normale Sup\'erieure,
and the Laboratoire de Physique Th\'eorique et Hautes Energies,
CNRS and Universit\'e Pierre et Marie Curie - Paris 6, and especially Constantin Bachas and
Jean-Bernard Zuber for their warm hospitality while part of this work was being completed.

%\newpage

\appendix

%%%%%%%%%%%%%%%%%%%%%%%%%%%%%%%%%%%%%%%%%%%
%%%%%%%%%%%%%%%%%%%%%%%%%%%%%%%%%%%%%%%%%%%
\section{Linearization around near-horizon Lifshitz}
\setcounter{equation}{0}
\label{AppA}
%%%%%%%%%%%%%%%%%%%%%%%%%%%%%%%%%%%%%%%%%%%
%%%%%%%%%%%%%%%%%%%%%%%%%%%%%%%%%%%%%%%%%%%

The near-horizon Lifshitz geometry present for $k=k_c$ smoothly extends into a solution
of the full non-linear reduced field equations which is asymptotic to AdS$_6$ for large $r$.
To date, these full non-linear equations have not been solved analytically, but the numerical
evidence for their existence is overwhelming. The numerical solution is obtained by solving
the full non-linear reduced equations starting from the near-horizon Lifshitz solution plus
first order perturbative corrections around Lifshitz. As is familiar from the study of the
purely magnetic brane in AdS$_5$, the first order fluctuations, subject to appropriate near-horizon
boundary conditions, will act as seeds for the full non-linear solution with the same boundary conditions.

\sm

In this Appendix, we shall compute the first order fluctuations around the Lifshitz solution for $k=k_c$.
To set up the problem, it is preferable to use the parameter $\a = z^{-1}$ instead of the dynamical
scaling exponent $z$. In terms of $\alpha$, the data of the Lifshitz solution may be expressed
as follows,
\bea
\label{A1}
U(r) = u_0 r^2 \hskip 0.2in & \hskip 1in &  E(r) = q_0
\no \\
V_1(r) = \alpha \ln (r) \, &&  G(r) = g_0 r^{2 \a -1}
\no \\
V_2 (r) = 0 \hskip 0.4in &&
\eea
Here, $u_0, q_0, g_0$, and $\alpha$ are constants related to $k$ and the value $b_0$
of the field $B$ near the horizon with the normalization $V_2(0)=0$. The values of the remaining
constants may be conveniently parameterized in terms of $\alpha$,
\bea
\label{A2}
b_0 = { \sqrt{30 \alpha} \over 2 \alpha +1} ~~~~ & \hskip 1in & q_0 =  { \sqrt{10 (1- \alpha)} \over 2 \alpha +1}
\no \\
u_0 =   { 20 \over (2 \alpha +1)^2} &&  g_0 =  - \sqrt{\a (1-\a)}
\eea
Reality of $b_0, q_0$ and $g_0$ clearly requires that $\a$ be restricted to the range,
\bea
\label{A3}
0 \leq \a \leq 1
\eea
This condition restricts $b_0^2$  to the interval $0 \leq b_0^2 \leq 15/4$, the upper value being
attained when $\a = 1/2$. It also restricts $q_0^2$ to the interval $0 \leq q_0^2 \leq 10$, the upper
value corresponding to the Reissner-Nordstrom black brane. Recall that the critical curve in terms
of near-horizon data $(b_0,q_0)$ is obtained by eliminating $\alpha$ from the above relations,
and takes the  form,
\bea
\label{A4}
3 (b_0^2 + q_0^2)^2 - 10 (b_0^2 + 3 q_0^2) =0
\label{b0q0crit}
\eea

\subsection{Linearized reduced field equations}

The linearization problem is parameterized as follows,
\bea
\label{A5}
U(r) & = & u_0 r^2 +  u_1(r)
\no \\
V_1(r) & = & \alpha \ln (r) +  v_1 (r)
\no \\
V_2 (r) & = &   v_2(r)
\no \\
E(r) & = & q_0 \Big ( 1 +  e_1(r) \Big )
\no \\
G(r) & = & g_0 r^{2 \a -1} \left ( 1 + { 1  \over \a} g_1(r) \right )
\eea
Here the fluctuation fields $u_1, v_1, v_2, e_1$, and $g_1$ are treated as first order quantities,
higher orders therein being neglected in the range where all these fields are small.
The precise normalization of $e_1$ and $g_1$ has been introduced for later convenience.

\sm

The corresponding linearized reduced Einstein field equations are given by,
\bea
\label{A6}
E1 \qquad
0 &= & r^2 v_1'' + r^2 v_2'' +2 \a r v_1' - 4 \a (1-\a) v_1 -2(1-\a)g_1
\no \\
E2 \qquad
0 & = & 20 r^2 v_1'' + 40 (2 \a+1) r v_1' + 40 \a r v_2' + \a (2\a+1)^2 r u_1'
-80\a (1-\a) v_1
\no \\ &&
+60 \a v_2 +(4 \a^4 + 12 \a^3 + 9 \a^2 +2 \a)u_1
+10(1-\a) e_1 -40 (1- \a) g_1
\no \\
E3 \qquad
0 & = & 20 r^2 v_2'' +40(1+\a) r v_2' +80\a(1-\a) v_1 -180 \a v_2
\no \\ &&
+(4 \a^4- 3 \a^2 -\a) u_1 +40(1-\a) g_1+10(1-\a) e_1
\no \\
E4 \qquad
0 & = & (2\a+1)^2 r^2 u_1'' + (8 \a^3+ 24 \a^2 + 18 \a +4) r u_1'
\no \\ &&
+ 80 r (v_1'+v_2') + (8 \a^4 +16 \a^3 + 18 \a^2 +10 \a+2) u_1
\no \\ &&
+ 160 \a (1-\a) v_1 +120 \a v_2 -60(1-\a) e_1 + 80 (1-\a) g_1
\eea
The reduced Maxwell-two-form equations are given by,
\bea
\label{A7}
M1 \qquad
0 & = & re_1' + 2 r (v_1' +  v_2') + 4 \a (v_1 + v_2) + 2 \a e_1 + 2 g_1
\no \\
M2 \qquad
0 & = & 20 r g_1' - \a (2 \a +1)^2 u_1' +40 \a r (v_1'- v_2')
\no \\ &&
-(4 \a^3 + 4 \a^2 + \a) u_1 + 40 \a (v_1-v_2) + 20 \a e_1 + 20 g_1
\eea
There is also the constraint,
\bea
\label{A8}
CON \qquad
0 & = & \a (2\a+1)^2 r u_1' + 40 (1+\a)r v_1' + 40 (2\a +1) r v_2'
\no \\ &&
+ (8 \a^4 + 12 \a^3 + 6 \a^2 + \a) u_1 + 80\a(1-\a) v_1 -120 \a v_2
\no \\ &&
+ 20 (1 -\a) e_1 + 40 (1-\a) g_1
\eea
whose $r$-derivative  is a linear combination of $E1, E2, E3, E4, M1$, and $M2$.

\subsection{Solution to the linearized equations}

As a result of the scale invariance of the Lifshitz background solution,
the linearized equations are invariant under an arbitrary rescaling of $r$.
This is manifest also from the observation that the equations above
are linear differential equations with constant coefficients with respect
to the differential operator $r\, d/dr$. Thus, the system may be solved
by a linear superposition of powers of $r$. Concretely, we set
\bea
\label{A9}
\left ( \matrix{ v_1 (r) \cr v_2 (r) \cr u_1 (r) \cr e_1(r) \cr g_1(r) \cr} \right )
= r^\lambda \cV
\hskip 1in
\cV = \left ( \matrix{ \cV_1 \cr \cV_2 \cr \cU_1 \cr \cE_1 \cr \cG_1 \cr} \right )
\eea
where $\cV_1, \cV_2, \cU_1, \cE_1, \cG_1$ are independent of $r$. The allowed values
of the exponent $\lambda $ are to be  determined by substituting (\ref{A9})
into equations (\ref{A6}), (\ref{A7}) and (\ref{A8}). The resulting equation may
be recast in the  form $\cA \cV =0$.
The explicit form of $\cA$ may be deduced from (\ref{A6}), (\ref{A7}) and (\ref{A8}),
but will not be written out here. Using Maple, one computes,
\bea
\label{A11}
\det \cA = 400 \a (2 \a +1)^2 \l^2 (\l+1) (\l + 2 \a +1)^2
\left ( \l^2 + (2 \a+1) \l - 6 \a^2 - \a -2 \right )
\eea
For the double zeros of the determinant, at $\lambda =0$ and at
$\lambda =-1-2\alpha$, the general solution is obtained by replacing $r^\lambda \cV$ in (\ref{A9})
by $ r^\lambda \cV  + r^\lambda \ln (r) \tilde \cV$.
Here, $\tilde \cV$ and $\cV$ obey the relations $\cA \tilde \cV=0$
and $\cA \cV + \cB \tilde \cV=0$, and we shall not need the explicit
form of $\cB$.

\sm

The modes corresponding to $\lambda =0$ and $\lambda =-1$ respectively correspond to
the dilation zero mode of the Lifshitz background, and to its translation mode. The dilation modes
are non-vanishing at the horizon, therefore modify the horizon data, and are thus excluded from
contributing. The $\lambda =-1$ and $\lambda =-1-2 \alpha$ modes are both singular at the
horizon, and cannot contribute either. This leaves the modes for which $\lambda$ obeys,
\bea
 \l^2 + (2 \a+1) \l - 6 \a^2 - \a -2=0
 \eea
This equation is solved as follows,
\bea
\lambda_\pm = - \a - \half \pm \Delta
\hskip 1in
\Delta = \half \sqrt{ 28 \a^2 + 8 \a +9}
\eea
In the interval $0 \leq \a \leq 1$, only the branch $\l_+$ is regular, since
we have $\l_- <-2$, and $\l_+ >1/2 $. Thus, only the branch $\l_+$
needs to be retained for matching. The corresponding eigenvector is
given by the following components,
\bea
\cV_1 & = & -\cJ ( 456 \a^5 + 164 \a^4 + 78 \a^3 -185 \a^2 -64 \a -44)
\no \\
\cV_2 & = &  \cJ ( 192 \a^5 + 376 \a^4 + 128 \a^3 + 106 \a^2 + 8)
%\no \\ &&
+ \Delta \cJ ( 12 \a^4 - 8 \a^3 + 49 \a^2 +16 \a +12)
\no \\
\cU_1 & = &  { 40 \cJ \over \a (2 \a +1)^2} \bigg (
192 \a^6 -560 \a^5 + 84 \a^4 -188 \a^3 + 75 \a^2 -20 \a +12
\no \\ && \hskip 1.2in
+ \Delta \{ 12 \a^5 + 76 \a^4 -47 \a^3 -21 \a^2 - 16 \a -4 \} \bigg )
\no \\
\cE_1 & = &  {4 \cJ \over 2 \a +1} \bigg ( 216 \a^5 -324 \a^4 - 70 \a^3 -183 \a^2 -24 \a -20
\no \\ && \hskip 1.2in
+ 2\Delta \{ -36 \a^4 + 12 \a^3 -17 \a^2 -4 \} \bigg )
\no \\
\cG_1 & = &  2 \cJ \left ( 840 \a^6 + 244 \a^5 + 286 \a^4 -149 \a^3 + 12 \a^2 -30 \a +12 \right )
\no \\ &&
+ 4 \Delta J \left ( 12 \a^5 + 28 \a^4 + 21 \a^3 + 3 \a^2 + \a -2 \right )
\eea
where $\cJ$ is the overall normalization of the eigenvector.

\sm

Maple calculations, using the above regular branch of the perturbative solution as initial
conditions, clearly demonstrate that the perturbative solution continues into a regular asymptotically
AdS$_6$ solution.
%The file is {\tt k=k-cMatching.mw}.

%%%%%%%%%%%%%%%%%%%%%%%%%%%%%%%%%%%%%%%%%%%
%%%%%%%%%%%%%%%%%%%%%%%%%%%%%%%%%%%%%%%%%%%
\section{Identification of $(b_0,q_0)$ in low $\hat T$ solutions at $k=k_c$}
\setcounter{equation}{0}
\label{AppB}
%%%%%%%%%%%%%%%%%%%%%%%%%%%%%%%%%%%%%%%%%%%
%%%%%%%%%%%%%%%%%%%%%%%%%%%%%%%%%%%%%%%%%%%

\begin{figure}[htb]
\centering
\begin{minipage}[c]{0.5\linewidth}
\centering \includegraphics[width=3.0in]{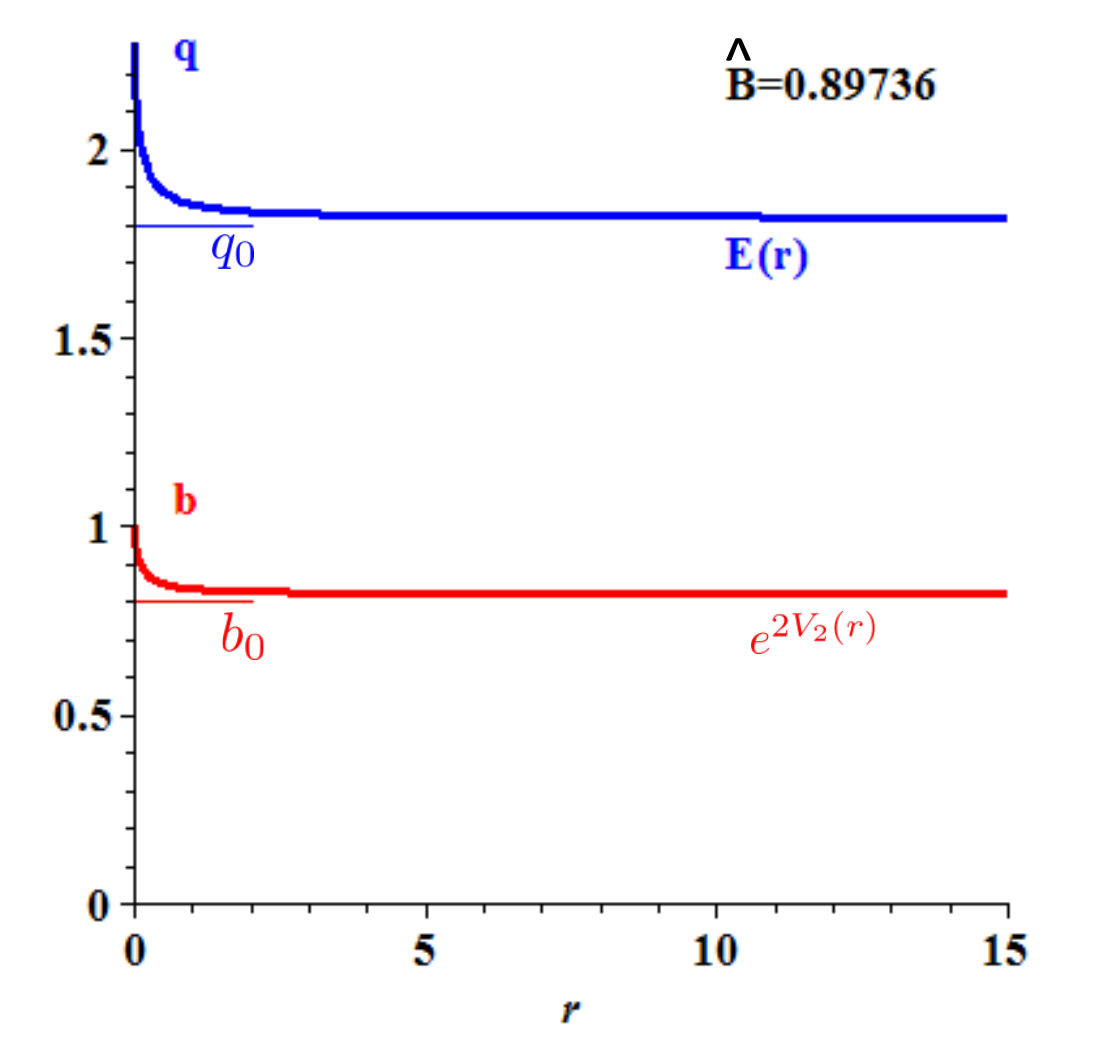}
\end{minipage}%
\begin{minipage}[c]{0.5\linewidth}
\centering \includegraphics[width=3.0in]{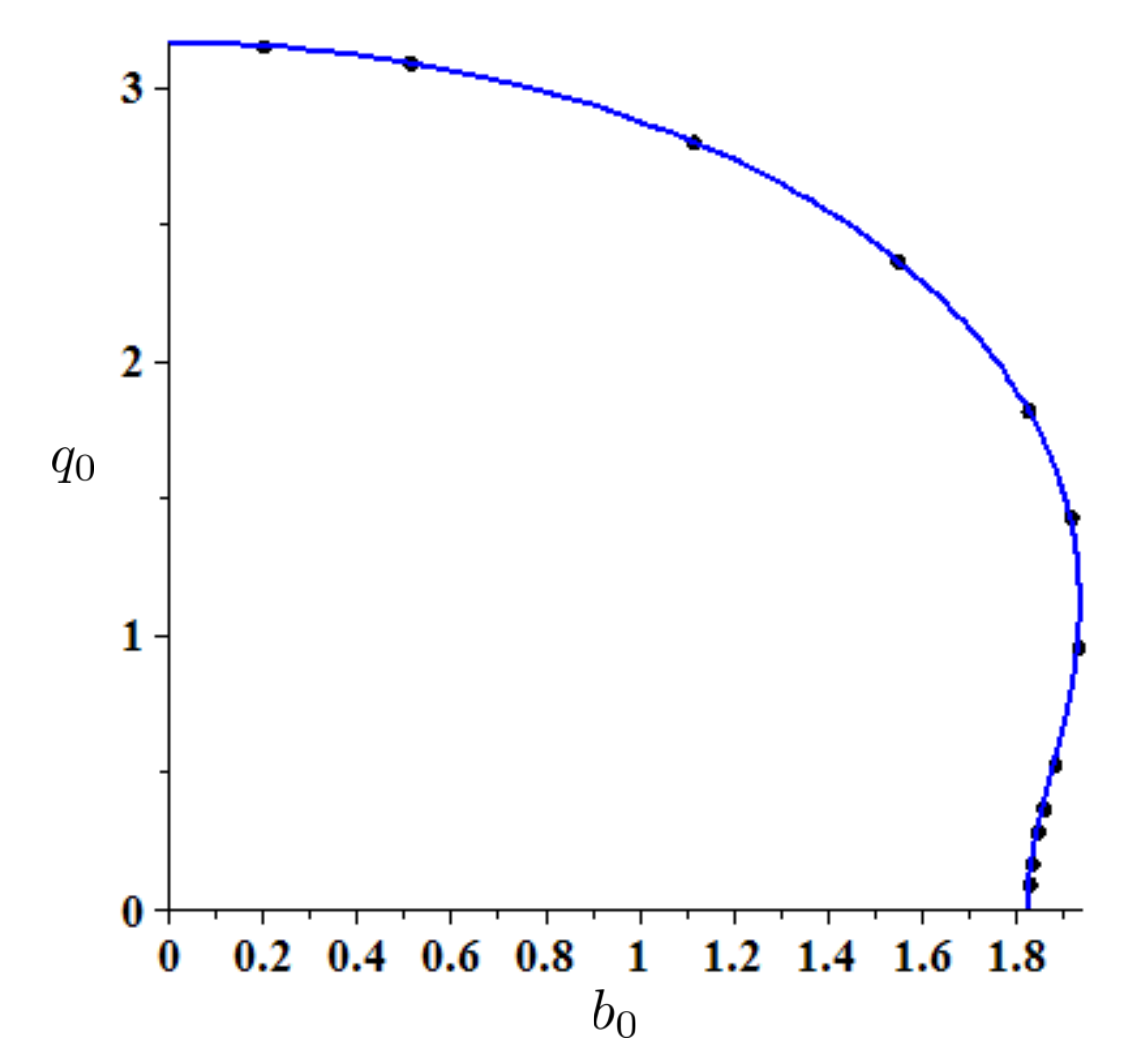}
\end{minipage}
\caption{Left panel: finite renormalization effects occur in transitioning between the
finite $\hat T$ initial data $(b,q)$ at the horizon,  and the initial data $(b_0,q_0)$ at the horizon of the
scaling solution. Right panel: the blue line represents the critical curve $3(b_0^2+e_0^2)^2=10b_0^2+30e_0^2$;
the data point were collected in the scaling region from the low $\hat T$ numerical solutions.}
\label{fig8}
\end{figure}

Numerical solutions, obtained with initial conditions at the horizon at finite temperature,
are governed by the data $q$ and $b$ at the horizon, as defined in the coordinate system employed in the numerics.  These parameters do not coincide
exactly with the parameters $q_0 $ and $b_0$ of the $\hat T=0$ solutions because
a finite renormalization effect from finite $\hat T$ boundary conditions to the $\hat T=0$
scaling solution takes place. This effect is illustrated in the left panel of  figure \ref{fig8}. At very low $\hat T$,
the functions $e^{2V_2(r)}$ and $E(r)$ settle to a value which is constant over a large
number of $e$-folds, and  to high precision. It is the values of $e^{2V_2(r)}$ and $E(r)$
in this regime that determine $b_0$ and $q_0$ in terms of $b$ and $q$.
%
%\begin{figure}[htb]
%\begin{centering}
%\includegraphics[scale=0.85]{AsymB_fig8}
%\caption{Finite renormalization effects occur in transitioning %between the
%finite $\hat T$ initial data at the horizon and the scaling %olution.}
%\label{fig8}
%\end{centering}
%\end{figure}
%

%

The corresponding assignments for $q_0$ and $b_0$, derived from the numerical runs
at finite but very low $\hat T$, may be plotted and compared with the critical curve
which is given  exactly by \ref{b0q0crit}. The corresponding plot
is given in the right panel of figure \ref{fig8}, showing excellent agreement throughout the curve.
%
%\begin{figure}[htb]
%\begin{centering}
%\includegraphics[scale=0.8]{AsymB_fig7}
%\caption{The blue line represents the critical curve %$3(b_0^2+e_0^2)^2=10b_0^2+30e_0^2$;
%the data point were collected in the scaling region from the low %$\hat T$ numerical solutions.}
%\label{fig7}
%\end{centering}
%\end{figure}
%

\end{document}